\documentclass[preprint]{emulateapj}
\usepackage{multirow}
\usepackage{natbib}
\usepackage{xspace}
\usepackage[colorlinks=false,linkcolor=black, citecolor=green, linktoc=page, urlcolor=cyan, 
pdffitwindow=false]{hyperref}
\usepackage[latin1]{inputenc}
\usepackage{amsmath}
\usepackage[usenames,dvipsnames]{color}

\def\reff@jnl#1{{\rm#1\/}}
\def\aj{\reff@jnl{AJ}}         
\def\araa{\reff@jnl{ARA\&A}}      
\def\apj{\reff@jnl{ApJ}}        
\def\apjl{\reff@jnl{ApJ}}        
\def\apjs{\reff@jnl{ApJS}}       
\def\aap{\reff@jnl{A\&A}}        
\def\aapr{\reff@jnl{A\&A~Rev.}}     
\def\aaps{\reff@jnl{A\&AS}}       
\def\mnras{\reff@jnl{MNRAS}}      
\def\physrep{\reff@jnl{Physics Reports}}
\def\prd{\reff@jnl{Phys.Rev.D}}     
\def\prl{\reff@jnl{Phys.Rev.Lett}}   
\def\pasp{\reff@jnl{PASP}}       
\def\pasj{\reff@jnl{PASJ}}       
\def\nat{\reff@jnl{Nature}}       
\def\jcap{\reff@jnl{JCAP}}   
\def\memsai{\reff@jnl{MemSAI}} 
\def\na{\reff@jnl{New Astronomy}}       

\def\Sref#1{$\S$\ref{#1}\xspace}

\def\Fref#1{Figure~\ref{#1}\xspace}

\def\Tref#1{Table~\ref{#1}\xspace}

\def\Eref#1{Equation~\ref{#1}\xspace}

\def\Aref#1{Appendix~\ref{#1}\xspace}
\def\Cref#1{Chapter~\ref{#1}\xspace}

\newcommand{\classstar}{{\tt CLASS\_STAR}}
\newcommand{\spreadmodel}{{\tt SPREAD\_MODEL}}
\newcommand{\modestclass}{{\tt MODEST\_CLASS}}
\newcommand{\magauto}{{\tt MAG\_AUTO}}
\newcommand{\magmodel}{{\tt MAG\_MODEL}}
\newcommand{\magdetmodel}{{\tt MAG\_DETMODEL}}
\newcommand{\magaper}{{\tt MAG\_APER\_4}}
\newcommand{\magerrauto}{{\tt MAGERR\_AUTO}}
\newcommand{\magerrmodel}{{\tt MAGERR\_MODEL}}

\newcommand{\magerraper}{{\tt MAGERR\_APER\_4}}
\newcommand{\flag}{{\tt FLAGS}}
\newcommand{\fluxradius}{{\tt FLUX\_RADIUS}}

\newcommand{\chihway}[1]{\textcolor{black}{ #1}}

\begin{document} 

\title[Transfer function for DES]{Modelling the Transfer Function for the Dark Energy Survey}

\author{C.~Chang$^{*}$\altaffilmark{1}, M. T.~Busha\altaffilmark{2,3}, R. H.~Wechsler\altaffilmark{2,3,4}, A.~Refregier\altaffilmark{1}, A.~Amara\altaffilmark{1}, E.~Rykoff\altaffilmark{2,3}, \\ 
M. R.~Becker\altaffilmark{2,3}, C.~Bruderer\altaffilmark{1}, L.~Gamper\altaffilmark{1}, B.~Leistedt\altaffilmark{5}, H.~Peiris\altaffilmark{5}, T.~Abbott\altaffilmark{6}, F. B.~Abdalla\altaffilmark{5,7}, \\ 
E. ~Balbinot\altaffilmark{8}, M.~Banerji\altaffilmark{9,10}, R. A.~Bernstein\altaffilmark{11}, E.~Bertin\altaffilmark{12}, D.~Brooks\altaffilmark{5}, A.~Carnero Rosell\altaffilmark{13, 14}, \\ 
S.~Desai\altaffilmark{15,16}, L.~N. da Costa\altaffilmark{13,14}, C.~E Cunha\altaffilmark{17}, T.~Eifler\altaffilmark{18}, A.E.~Evrard\altaffilmark{12,19,20}, A.~Fausti Neto\altaffilmark{14}, \\ 
D.~Gerdes\altaffilmark{19}, D.~Gruen\altaffilmark{21, 22}, D.~James\altaffilmark{6}, K.~Kuehn\altaffilmark{23}, M. A. G.~Maia\altaffilmark{13,14}, M.~Makler\altaffilmark{24}, \\ 
R.~Ogando\altaffilmark{13, 14}, A.~Plazas\altaffilmark{25}, E.~Sanchez\altaffilmark{26}, B.~Santiago\altaffilmark{27, 14}, M.~Schubnell\altaffilmark{19}, I.~Sevilla-Noarbe\altaffilmark{26}, \\ 
C.~Smith\altaffilmark{6}, M.~Soares-Santos\altaffilmark{28}, E.~Suchyta\altaffilmark{29}, M. E. C.~Swanson\altaffilmark{30}, G.~Tarle\altaffilmark{19}, J.~Zuntz\altaffilmark{31}}
\altaffiltext{1}{Department of Physics, ETH Zurich, Wolfgang-Pauli-Strasse 16, CH-8093 Zurich, Switzerland}
\altaffiltext{2}{Kavli Institute for Particle Astrophysics and Cosmology, P.O. Box 2450, Stanford, CA 94305, USA}
\altaffiltext{3}{SLAC National Accelerator Laboratory, 2575 Sand Hill Road, Menlo Park, CA 94025, USA}
\altaffiltext{4}{Department of Physics, Stanford University, 382 Via Pueblo Mall, Stanford, CA 94305, USA}
\altaffiltext{5}{Department of Physics and Astronomy, University College London, London WC1E 6BT, UK}
\altaffiltext{6}{Cerro Tololo Inter-American Observatory, National Optical Astronomy Observatory, Casilla 603, La Serena, Chile}
\altaffiltext{7}{Department of Physics and Electronics, Rhodes University, PO Box 94, Grahamstown, 6140 South Africa}
\altaffiltext{8}{Department of Physics, University of Surrey, Guildford GU2 7XH, UK}
\altaffiltext{9}{Institute of Astronomy, University of Cambridge, Madingley Road, Cambridge CB3 0HA, UK}
\altaffiltext{10}{Kavli Institute for Cosmology, University of Cambridge, Madingley Road, Cambridge CB3 0HA, UK}
\altaffiltext{11}{Carnegie Observatories, 813 Santa Barbara St., Pasadena, CA 91101, USA}
\altaffiltext{12}{Institut d'Astrophysique de Paris, Univ. Pierre et Marie Curie \& CNRS UMR7095, F-75014 Paris, France}
\altaffiltext{13}{Observat\'orio Nacional, Rua Gal. Jos\'e Cristino 77, Rio de Janeiro, RJ - 20921-400, Brazil}
\altaffiltext{14}{Laborat\'orio Interinstitucional de e-Astronomia - LIneA, Rua Gal. Jos\'e Cristino 77, Rio de Janeiro, RJ - 20921-400, Brazil}
\altaffiltext{15}{Department of Physics, Ludwig-Maximilians-Universit\"{a}t, Scheinerstr.\ 1, 81679 Munich, Germany}
\altaffiltext{16}{Excellence Cluster Universe, Boltzmannstr.\ 2, 85748 Garching, Germany}
\altaffiltext{17}{Robert Bosch LLC, 4009 Miranda Ave, Suite 225, Palo Alto, CA 94304, USA}
\altaffiltext{18}{Jet Propulsion Laboratory, California Institute of Technology, 4800 Oak Grove Dr., Pasadena, CA 91109, USA}
\altaffiltext{19}{Department of Physics, University of Michigan, Ann Arbor, MI 48109, USA}
\altaffiltext{20}{Department of Astronomy, University of Michigan, Ann Arbor, MI 48109, USA}
\altaffiltext{21}{University Observatory Munich, Scheinerstrasse 1, 81679 Munich, Germany}
\altaffiltext{22}{Max Planck Institute for Extraterrestrial Physics, Giessenbachstrasse, 85748 Garching, Germany}
\altaffiltext{23}{Australian Astronomical Observatory, North Ryde, NSW 2113, Australia}
\altaffiltext{24}{ICRA, Centro Brasileiro de Pesquisas F\'isicas, Rua Dr. Xavier Sigaud 150, CEP 22290-180, Rio de Janeiro, RJ, Brazil}
\altaffiltext{25}{Brookhaven National Laboratory, Bldg 510, Upton, NY 11973, USA}
\altaffiltext{26}{Centro de Investigaciones Energ\'eticas, Medioambientales y Tecnol\'ogicas (CIEMAT), Madrid, Spain}
\altaffiltext{27}{Instituto de F\'\i sica, Universidade Federal do Rio Grande do Sul, Av. Bento Gon\c calves, 9500, Porto Alegre, RS - 91501-970, Brazil}
\altaffiltext{28}{Fermi National Accelerator Laboratory, P. O. Box 500, Batavia, IL 60510, USA}
\altaffiltext{29}{Center for Cosmology and Astro-Particle Physics, The Ohio State University, Columbus, OH 43210, USA}
\altaffiltext{20}{National Center for Supercomputing Applications, 1205 West Clark St., Urbana, IL 61801, USA}
\altaffiltext{31}{Jodrell Bank Center for Astrophysics, School of Physics and Astronomy, University of Manchester, Oxford Road, Manchester, M13 9PL, UK}

\begin{abstract}
  We present a forward-modelling simulation framework designed to model the data products from the Dark 
  Energy Survey (DES). This forward-model process can be thought of as a transfer function --- a mapping 
  from cosmological/astronomical signals to the final data products used by the scientists. Using output from 
  the cosmological simulations (the Blind Cosmology Challenge), we generate simulated images 
  \citep[the Ultra Fast Image Simulator,][]{2013AC.....1...23B} and catalogs representative of the DES data.  
  In this work we demonstrate the framework by simulating the 244 deg$^{2}$ coadd images and catalogs in 
  5 bands for the DES Science Verification (SV) data. The simulation output is compared with the corresponding 
  data to show that major characteristics of the images and catalogs can be captured. We also point out several 
  directions of future improvements. Two practical examples -- star-galaxy classification and proximity effects on 
  object detection -- are then used to illustrate how one can use the simulations to address systematics issues in 
  data analysis. With clear understanding of the simplifications in our model, we show that one can use the 
  simulations side-by-side with data products to interpret the measurements. This forward modelling approach is 
  generally applicable for other upcoming and future surveys. It provides a powerful tool for systematics studies 
  which is sufficiently realistic and highly controllable. 

\end{abstract}


\section{Introduction}
\label{sec:intro}
\setcounter{footnote}{0}

We have entered an exciting era of optical surveys. In recent years, the Kilo Degree 
Survey\footnote{\url{http://kids.strw.leidenuniv.nl/}} \citep[KiDS,][]{2013ExA....35...25D}, 
the Panoramic Survey Telescope and Rapid Response System\footnote{\url{http://pan-starrs.ifa.hawaii.edu/public/}} 
\citep[Pan-STARRS,][]{2004AN....325..636H}, the 
Hyper Suprime-Cam Survey\footnote{\url{http://www.naoj.org/Projects/HSC/}} \citep[HSC,][]{2012SPIE.8446E..0ZM}, 
and the Dark Energy Survey\footnote{\url{http://www.darkenergysurvey.org/}} \citep[DES,][]{2005astro.ph.10346T} have 
all started to take data. In particular, DES will cover the widest area (one eighth of the sky), and the resulting enormous 
datasets will allow one to achieve very high statistical precision in measuring cosmological parameters. We will soon 
be able to test with multiple cosmological probes, the standard $\Lambda$CDM cosmological model, and gain a 
better understanding of the nature of Dark Energy \chihway{\citep{2006astro.ph..9591A, 2008ARAA..46..385F, 
2010GReGr..42.2177H, 2011ARAA..49..409A, 2013PhR...530...87W, 2014daen.book.....R}.}

As the statistical uncertainties are reduced by orders of magnitude in these large datasets, various systematic 
uncertainties in analysing the data become important \chihway{\citep{2006MNRAS.366..101H, 2008MNRAS.391..228A, 
2013arXiv1311.2597H, 2014JCAP...04..007A, 2014ApJ...795...45S}.} 
Different cosmological probes are sensitive to different systematic effects. But generally, as all 
measurements begin from the same processed images and catalogs, the first-order systematic effects in these 
data products need to be well understood. In other words, one needs to understand how the 
information coming from the sky is transformed into the processed images and catalogs on which we base our 
scientific measurements. Moreover, one needs to understand how this transformation depends on the properties 
of the astronomical sources and the observing conditions. This paper seeks to understand this complicated 
process -- the ``transfer function'' -- for DES via forward-modelling. \chihway{The goal of this work is to model the 
coadd images and the catalogs from DES. Although this framework still contains several simplifications 
(see \Sref{sec:simplification}), it is the necessary first step in building a fully realistic simulation pipeline.}
Note also that although we focus on DES in this paper, our methodology is generally applicable for all upcoming 
and future large surveys.

\begin{figure*}
  \begin{center}
  \includegraphics[scale=0.4]{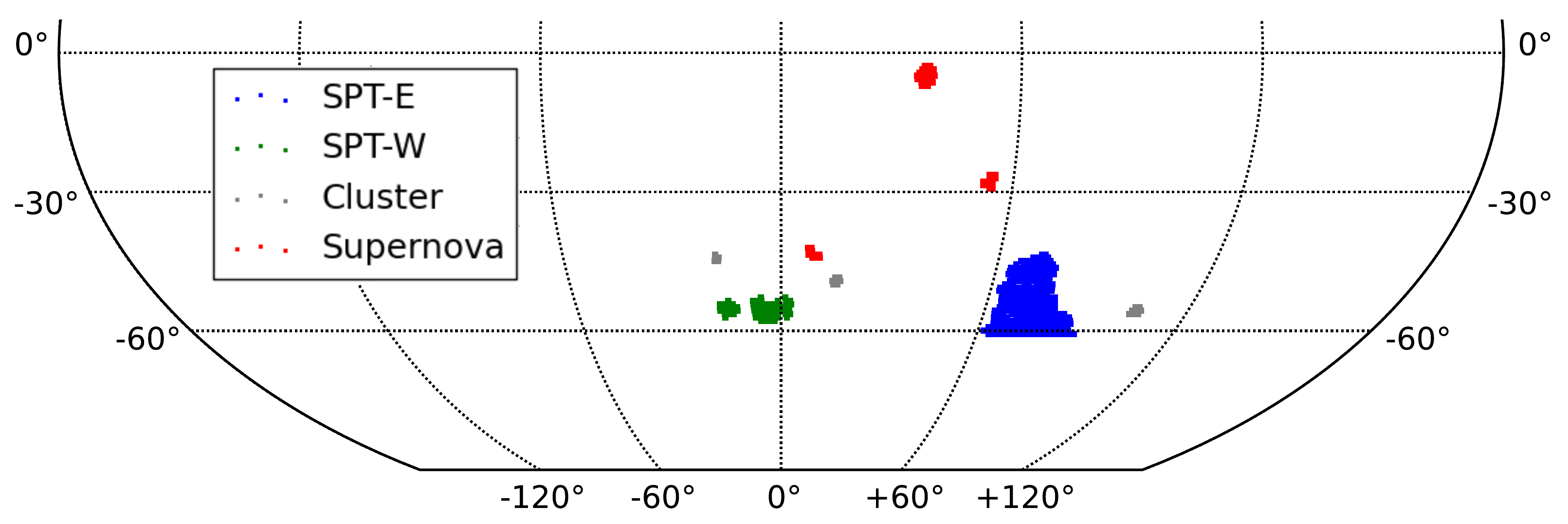}  
  \vspace{-0.1in}
  \end{center}
  \caption{Footprint for the DES SV data used in this work. The different colours indicate the 
    different types of fields: the blue and green areas are the SPT wide-field coverage, the grey areas 
    indicate the pointed cluster fields outside of the SPT fields, and the red areas indicate the Supernova
  fields.}
\label{fig:footprint}
\end{figure*}

The concept of modelling the transfer function for a specific experiment has a long history in the field of particle 
physics \chihway{\citep{1987CoPhC..46...43B, 1990sna..conf...12N, 1992CoPhC..67..465M, 2003NIMPA.506..250A, 
2010mcss.book.....B, 2012PhRvD..86a0001B}.} 
In fact, the results of particle physics experiments can only 
be interpreted in terms of their corresponding Monte Carlo simulations. In optical astronomy, however, the idea of 
forward-modelling is less mature, despite the fact that highly developed simulation tools exist for individual steps of 
the transfer function. For example, cosmological simulations such as \citet{2009AA...499...31H, 
2011MNRAS.414.2235K, 2013ApJS..208....1G, 2013AN....334..691R, 2013MNRAS.tmp.2702W} begin with 
N-body simulations and develop prescriptions for assigning 
astronomical objects to dark matter halos. \citet{2003MNRAS.339..289S, 
2008MNRAS.385.1443S} and \citet{2012MNRAS.425.3024V} use different techniques to simulate various 
hydrodynamic processes in structure formation and link to observables related to cosmology. 
\chihway{
\citet{2002AJ....124..266P} uses simulated galaxy images to help understand the study of galaxy morphology. 
\citet{2009MmSAI..80..422B, 2010MNRAS.405.2044B, 2012MNRAS.423.3163K, 2013AC.....1...23B} simulate 
astronomical images with simple instrumental effects to understand how well one can recover information from 
noisy data. Finally, \citet{2013ascl.soft07011P} focuses on the detail modelling of the astronomical instrument 
to understand how the instrument design affects the imaging data. }Although 
these different simulations are very helpful for understanding the technical issues in the separate areas, one cannot 
straightforwardly infer how the results in different parts of the transfer function couple to each other. The recent 
attempt described in \citet{2010SPIE.7738E..53C} is one of the first efforts to consolidate the issue by connecting all 
types to an end-to-end simulation framework for one specific project, the Large Synoptics Survey Telescope (LSST). 
Our work is based on the same philosophy, but instead of modelling a future instrument like LSST, the aim is to 
model DES, which is currently taking data. 

We extend from the Blind Cosmology Challenge simulations \citep[BCC,][]{2013AAS...22134107B} 
to include processed images from the Ultra Fast Image Generator \citep[\textsc{UFig},][]{2013AC.....1...23B} and 
catalog products which come from a similar analysis pipeline as that used in the DES Data Management 
\citep[DESDM,][]{2006SPIE.6270E..23N, 2011arXiv1109.6741S, 2012ApJ...757...83D, 2012SPIE.8451E..0DM}. 
\chihway{
Our implementation is similar to the earlier DES data challenges described in \citet{2010AAS...21547007L} and 
\citet{2011arXiv1109.6741S}, where DES simulations were generated before the existence of data to test data 
management and science analysis software. This work is complementary to the earlier data challenges 
in that the simulations in this work is guided by the actual DES data and data processing pipeline being used, 
which was not available at the time of the data challenge. }

This paper is organised as follows: In \Sref{sec:data}, we briefly introduce the Dark Energy Survey and the relevant 
data products that are used in this paper. In \Sref{sec:simulations} we describe in detail the forward-modelling 
framework, including individual simulation and analysis tools, as well as the interfacing between them. A series of 
quality assurance tests are performed in \Sref{sec:qa} to examine the output products of our framework. We 
cross-check with early DES data to ensure the output captures the main characteristics of the data. We then 
demonstrate in \Sref{sec:applications} two practical applications where we use this forward-modelling framework 
to address specific technical questions in the data analysis process. Finally, we conclude in \Sref{sec:summary}.

An example of the simulation output and supporting documentation from this work can be found at 
\url{http://www.phys.ethz.ch/~ast/cosmo/bcc_ufig_public/}. 

\section{The Dark Energy Survey} 
\label{sec:data}

The Dark Energy Survey (DES) is a wide-field optical survey that officially began in August 2013 
\citep{2014SPIE.9149E..0VD} and will continue to 
survey the sky through 2018. The full DES footprint will cover one eighth of the full sky (5,000 deg$^{2}$) in five optical 
bands ($grizY$). The homogeneous wide-field nature of the dataset will be important for cosmology studies 
on very large scales. The primary instrument for DES is a newly assembled wide-field (3 deg$^{2}$) 
mosaic camera, the Dark Energy Camera \citep[DECam,][]{2012AAS...21941305D}, installed on the 4m Blanco 
telescope at the Cerro Tololo Inter-American Observatory (CTIO) in Chile. 

The raw images taken each night are collected and jointly processed with the DESDM 
software. In addition to the zeroth-order image processing (flat-fielding, bias correction, de-trending etc.), 
the DESDM pipeline contains mainly software packages described in \citet{2006SPIE.6270E..23N, 2011arXiv1109.6741S, 
2012ApJ...757...83D, 2012SPIE.8451E..0DM} -- \textsc{SCAMP} \citep[astrometry,][]{2006ASPC..351..112B}, 
\textsc{SWARP} \citep[image coaddition,][]{2002ASPC..281..228B}, \textsc{PSFEx} 
\citep[modelling of the point-spread-function,][]{2011ASPC..442..435B} and \textsc{SExtractor} 
\citep[object detection and measurement,][]{1996AAS..117..393B}. With continual improvement in the 
pipeline, DESDM performs regular releases of the data products. The main product from DESDM are 
images and catalogs of objects with calibrated properties.  

The initial pre-season of DES observations were labeled as Science Verification
(SV) imaging, which took place from November 2012 -- February 2013.  These
images were processed by the DESDM pipeline version
``SVA1'' (Yanny et al., in prep) to produce coadd images and \textsc{SExtractor}
catalogs. Additional quality checks and calibration were performed by DES
scientists, which included cropping out bad regions contaminated by satellite
and airplane trails, as well as the region at declination $<-61^\circ$ which has
a very high stellar density due to the presence of the Large Megallanic Cloud
(SVA1 Gold; Rykoff et al., in prep). After all cuts, the total sky coverage is 244 
$\mathrm{deg}^2$ of $griz$ imaging.  This includes several selected wide fields,
pointed cluster fields (RXC J2248.7-4431, 1E 0657-56, SCSO J233227-535827, 
and El Gordo), and deep supernova (SN) fields. \Fref{fig:footprint}
shows the full SVA1 footprint and how the different fields are distributed.
The SN fields are revisited every 5-7 days with longer exposures, and are
therefore 1-2 magnitudes deeper than the other fields, particularly in the $i$
and $z$ bands.  In this work, we base our forward-modelling framework on the
SVA1 Gold catalogs.  As the DESDM software and image quality continue to
improve for future releases, our modelling framework will adjust accordingly.

\section{Forward-modelling}
\label{sec:simulations}
\chihway{
In this section we briefly introduce the three major elements of our forward-modelling framework: two 
simulation tools (\Sref{sec:bcc}, \Sref{sec:ufig}) and the analysis software (\Sref{sec:desdm}). We then 
describe how the interfaces between the three components are implemented (\Sref{sec:bridge}) and 
the computational cost (\Sref{sec:computational}). First, however, we list in \Sref{sec:simplification} the 
main simplifications used in this framework. }

\subsection{Simplifications}
\label{sec:simplification}

The current framework as described below contains several simplifications. As we will discuss in 
\Sref{sec:summary}, more sophistication and realism is planned to be added to the framework as required 
from different science cases. The main simplifications of the current framework are the following: 
(1) We begin the forward-modelling from coadd images instead of single-exposure images, thus bypass 
the process of stacking images. (2) The PSF, airmass, background (limiting magnitude), quantum efficiency, 
and throughput are constant in each filter with no spatial variation across an image. (3) The background 
model is simplistic (Gaussian noise plus Lanczos resampling) and does not properly model the correlation 
of noise in the images. (4) There are no artefacts such as bad/hot columns on the detectors, satellites, 
cosmic rays, etc.. 

\chihway{It is important to stress that the focus of this forward-modelling framework is not to make 
simulations that are \textit{identical} to the data (nor is it possible to do so exactly). 
Rather, it is to capture the important characteristics of the data in a controlled environment where we know 
the truth. This allows us to interpret the measurements in a clean fashion within the limitations of the simulations.
 As a result, despite these simplifications, many data-related 
issues can already be investigated as we demonstrate in \Sref{sec:qa} and \Sref{sec:applications}. The 
results from these simplified simulations would also be important for interpreting more realistic simulations 
in the future as we incorporate more physics in the forward model. }

\subsection{The mock sky catalog}
\label{sec:bcc}

The primary input to our framework is a mock sky catalogs of astronomical sources. In this work, we use the 
Aadvark v1.0d catalogs generated as part of the BCC. The BCC catalog generation begins 
with particle light cones from a series of large (1-4 Gpc/h) N-body simulations with a defined cosmology (a 
flat LCDM cosmology in this case). The Adding Density Determined GAlaxies to Lightcone Simulations 
algorithm \citep[ADDGALS,][]{2013AAS...22134107B} associates galaxies to the dark matter particles by 
using a Sub-Halo Abundance Matching (SHAM) catalog \citep{2006ApJ...647..201C, 2010ApJ...717..379B} 
generated from a high resolution, low-volume tuning simulation to determine a probabilistic relation between 
a galaxy's magnitude and its local dark matter density. The algorithm then assigns basic properties (luminosity, 
colour, etc.) to each galaxy using a training 
set of spectroscopic data from the SDSS DR6 Value-Added Galaxy Catalog \citep{2005AJ....129.2562B} to 
match simulated galaxies to observed counterparts using the local galaxy environment. The training procedure 
is performed at low redshift and extrapolated to high redshift so that the colour 
distribution simultaneously matches the photometric data in SDSS DR8 and DEEP2. 
The intrinsic shape and size of each galaxy is then set to match to observations from the 
SuprimeCam deep $i'$-band data \citep{2012Natur.487..202D}. Finally, 
the galaxies are lensed by the multiple-plane ray-tracing code, Curved-sky grAvitational Lensing for 
Cosmological Light conE simulatioNS \citep[CALCLENS,][]{2013MNRAS.435..115B} to give perturbed 
shapes, positions and magnitudes. Additionally, a stellar distribution is added based on the TRIdimentional 
modeL of thE GALaxy code \citep[Trilegal,][]{2012rgps.book..165G, 2012ASPC..461..287B}, and the quasar 
model is based on \citet{2012MNRAS.424.2876M}. The full details of the BCC catalogs would be described in 
an upcoming paper.

These BCC catalogs serve as the ``true'' sky after the sources have been lensed by the large scale structures 
before the light enters the atmosphere. For this work, the main properties used in the BCC catalogs are the 
magnitude, size, colour, redshift and shape distributions of objects. The main requirement is that these 
distributions in the BCC catalog are modelled for objects fainter than the limiting magnitude of the dataset we wish 
to model. 

There are several advantages of using such sophisticated cosmological simulations as our input compared 
to using parametrised star/galaxy distributions [cf. our earlier work in 
\citet{2013AC.....1...23B}]. First, one preserves the cosmological clustering of 
the galaxies. Second, one simultaneously retains a self-consistent cosmology between clustering, lensing, 
and redshift evolution of galaxies. Finally, the correlation between the magnitudes of objects in different filter 
bands (i.e. colours) are also self-consistent. Note however, that the BCC catalogs cut off at a magnitude only 
slightly deeper than the DES main survey limiting magnitude. This suggests that the fainter objects that contribute 
to the background will be missing in our images and we cannot simulate properly the deeper Supernova fields. 
One would need to examine the impact of these missing faint objects on the measurement of interest when 
using the simulations from this framework. 
  
\subsection{The image simulation software}
\label{sec:ufig}

The Ultra Fast Image Generator \citep[\textsc{UFig}, for full detail of the implementation of \textsc{UFig}, see,][]{2013AC.....1...23B} 
is a fast image simulation code that generates scientific astronomical images that capture the major characteristics of 
a given instrument, as specified by the user. The computational time required for \textsc{UFig} to generate images in this work 
is much shorter than the time required to analyse the images (see \Sref{sec:computational}).

We briefly describe here the image rendering process in \textsc{UFig}. 
First, the apparent magnitudes of stars and galaxies are converted into number of photons 
expected at the focal plane, given the atmosphere and instrumental throughput in the specific filter band. 
Then, images of the galaxies are generated by drawing probabilistically, one photon at a time, from the galaxy profile 
model \citep[single S\'{e}rsic profile with varying S\'{e}rsic index,][]{1963BAAA....6...41S}. Next, we construct a model 
for the point spread function (PSF) given a desired seeing value. The galaxies are then convolved with the PSF model 
by displacing the photons randomly according to a probability density function described by the PSF profile. The image 
is then pixelated. Stars are generated directly on the pixels, with the same profile as the PSF model and appropriate 
Poisson noise on the pixel values. The stars and galaxies are generated via different approaches to optimise the 
computational speed. These pixel values are then converted into electronic units (ADUs) and an user-specified 
Gaussian noise is added. Finally, the full image is convolved with a Lanczos filter of size 3 \citep{Duchon:1979aa} to 
simulate the correlation of the noise in a coadd image. The full image is then rescaled to a given magnitude zeropoint.           

\subsection{The data processing software}
\label{sec:desdm}

As mentioned in \Sref{sec:data}, the DESDM pipeline uses a suite of software packages to produce the final catalog. 
Since we simulate the processed coadd images directly from \textsc{UFig} (\Sref{sec:ufig}), we bypass several steps in the 
DESDM pipeline. These are simplifications that can be improved upon in the future. The two \textit{main} packages 
involved in our framework are \textsc{PSFEx} and \textsc{SExtractor}.  

\textsc{PSFEx} is a software that constructs a model for the PSF of an image. Accurately knowing the PSF is important for 
later steps in the pipeline such as photometry measurements and galaxy profile-fitting. \textsc{SExtractor} is the main 
measurement software in the process. It estimates the background, detects objects, and conducts the basic 
measurements for each object. These include magnitudes estimated with several different approaches, various size 
estimates, parametrised model of the object profile, and classifiers that help the user identify different types of objects.
As the output is sensitive to detailed settings in the \textsc{PSFEx} and \textsc{SExtractor} configuration, we match the setting to 
that used in the SVA1 catalog whenever possible. 

\subsection{Bridging heaven and earth}
\label{sec:bridge}

The three basic elements of the forward-modelling framework described above are interfaced and connected as 
described in the following steps. 

\begin{figure*}
  \begin{center}
  \includegraphics[scale=0.22]{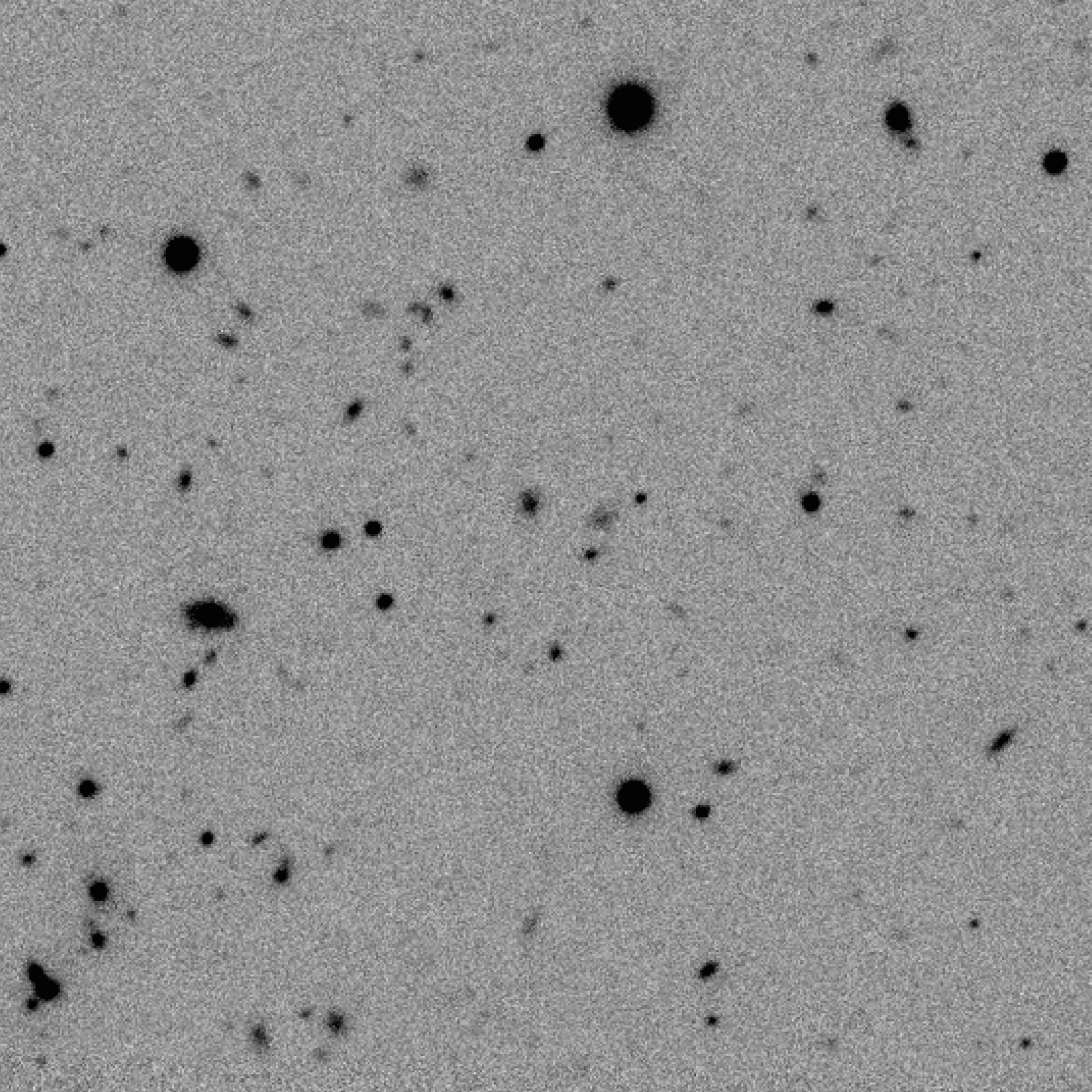} 
  \hspace{0.2in}
  \includegraphics[scale=0.22]{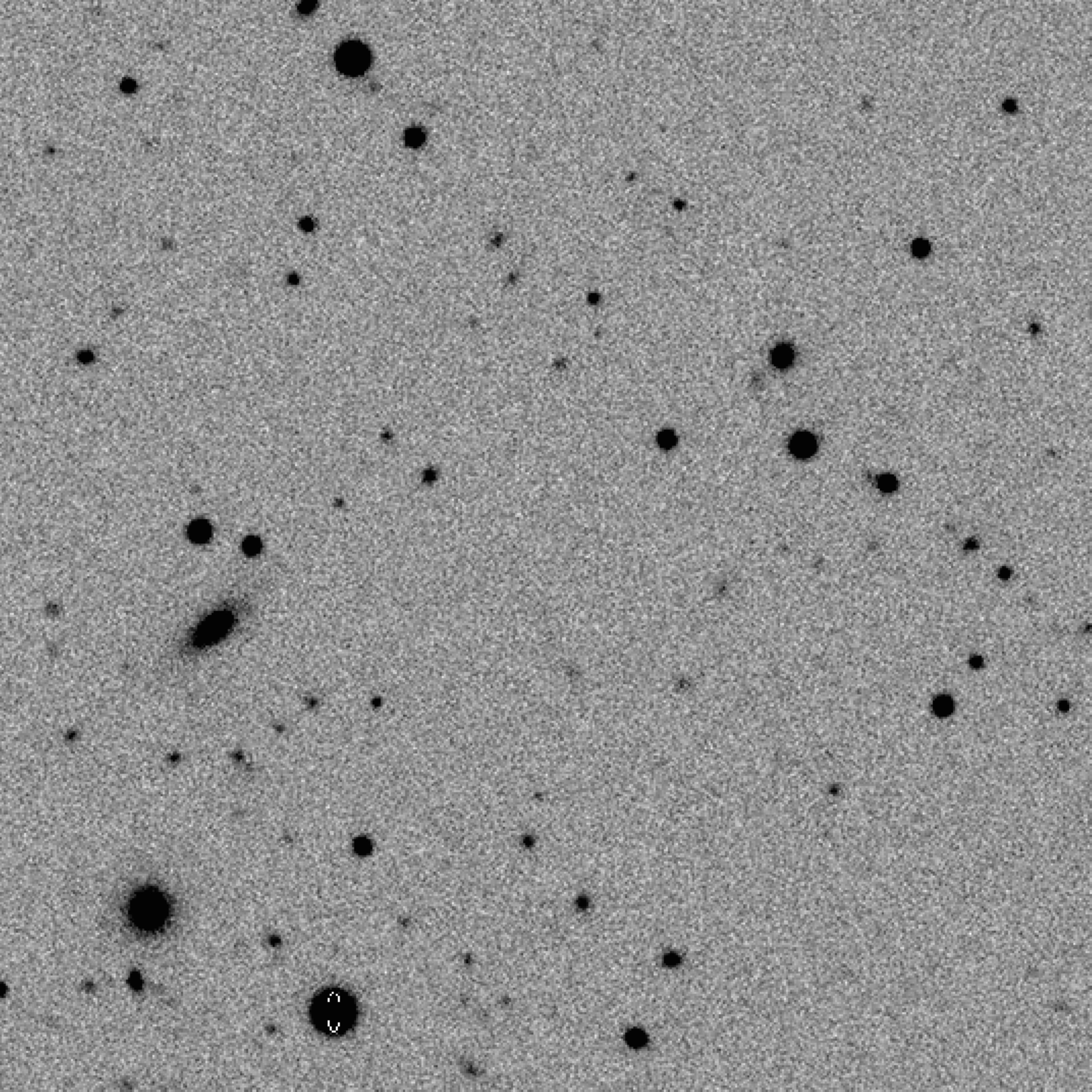}
  \end{center}
  \caption{A 500$\times$500 pixel region of an arbitrary $i$-band DES image (left) and its 
  simulation counterpart (right). The scales in both images are the same. Note that the objects are not matched 
  one-to-one in these images, but the statistical clustering and object properties appear qualitatively similar. 
  Note also that the texture of the background is slightly different in the simulations compared to the data, 
  indicating that improvements are needed for the background model.}
\label{fig:images}
\end{figure*}

\subsubsection{BCC catalog $\rightarrow$ \textsc{UFig} catalog}

The first step involves converting the ``sky information'' in the BCC catalogs into ``image information'' that can be 
used by \textsc{UFig}. We start by defining pointing positions on the sky, from which we draw a 0.75$\times$0.75 deg$^{2}$ 
area where the image will be simulated. The image size is defined by that of DESDM coadd images. 

The information in the BCC catalogs is then translated into \textsc{UFig} internal parameters. Object coordinates are 
converted into physical positions on the image with the appropriate World Coordinate System (WCS) 
transformation. All images are linearly projected from the sky with a pixel scale of 0.27 
arcsec/pixel\footnote{The measured pixel scale on the DES SV data is closer to 0.263 arcsec/pix. Changing 
the pixel scale by this amount (2.7\%) would however not result in the significant difference in our analysis.}. 
The apparent magnitude of stars and galaxies, as well as the ellipticity of galaxies are taken directly from the BCC 
catalogs. The intrinsic galaxy size information is based on the BCC catalogs but adjusted slightly so that the 2d 
distribution in apparent magnitude and intrinsic size is consistent with that derived from the COSMOS data 
\citep{2009AA...504..359J}. The adjustment is needed because the BCC catalog takes an approximate approach 
when converting the observed galaxy size into the intrinsic galaxy size. Finally, the galaxy is modelled by a single 
S\'{e}rsic profile, where the S\'{e}rsic indicies are band-independent and drawn randomly from the 
following distributions:  
\begin{equation}
f(n) = 0.2 \\
+ \left\{ 
\begin{array}{ll}
\exp(N(0.3,0.5) + N (1.6,0.4)) & \mbox{if $i < 20 $}; \\
\exp(N(0.2,1) ) & \mbox{if $i \geq 20 $}, 
\end{array} \right. 
\label{eq:sersic}
\end{equation}
$N(\mu, \sigma)$ denotes a normal distribution of mean $\mu$ and standard deviation $\sigma$. 
\Eref{eq:sersic} was derived in \citet{2013AC.....1...23B} from fitting deep $i$-band images 
\citep{2012ApJS..200....9G}. A more sophisticated S\'{e}rsic distribution that also takes into account 
the band dependencies would be a direction of future improvement. The S\'{e}rsic index is the only 
parameter of the source properties external to the BCC catalogs. 

\subsubsection{\textsc{UFig} catalog  $\rightarrow$ \textsc{UFig} image}

Next, we simulate a \textsc{UFig} image from the source catalog generated from the previous step. 
The instrument characteristics and observing conditions need to be specified for each image. These 
parameters include the throughput, the Charge-Coupled Device (CCD) characteristics, the seeing 
condition and the sky brightness. 

In all the simulations in this paper, we take the major instrumental parameters from the official DES 
Exposure Time 
Calculator\footnote{\url{http://www.ctio.noao.edu/noao/content/Exposure-Time-Calculator-ETC-0}} (ETC) 
as listed in \Tref{tab:ufig_params}. The atmospheric throughput describes the fraction of light 
that passes through the atmosphere at zenith. The telescope throughput describes the fraction of light 
that passes through the telescope and arrives at the focal plane. The mean wavelength and the bandwidth 
specify the basic properties of the filters. The quantum efficiency measures the fraction of photons that is 
converted into digital signal in the CCD. All quantities in this table are average values. Note also that we 
follow the DESDM convention and normalise the coadd images to either 90 ($griz$-band) or 45 ($Y$-band) 
seconds-equivalent exposures. 

On the other hand, the image-specific parameters (eg. exposure time, seeing, background noise) are 
tuned to the specific data we wish to model. \chihway{
We use a circular Moffat PSF model with $\beta=3.5$ \citep{1969AA.....3..455M}, 
which is is typically a good description for ground-based optical PSFs. The PSF is assumed to be spatially 
constant in each image and have a FWHM (which can be specified for a Moffat profile with given $\beta$ 
parameter) equal to the mean seeing in the data of interest. Similarly, the background level is set so 
that the expected limiting magnitude agrees with the data (see \Aref{sec:noise} for details on the derivation 
of the background noise).}

\Fref{fig:images} shows one arbitrary DES image in $i$-band and its simulation counterpart. Note that the 
objects in the images are not matched one-to-one, but the statistical clustering and noise properties appear 
qualitatively similar from visual inspection. We also note that due to the simplification in the background 
model (Gaussian noise plus Lanczos resampling), the texture of the background appears to be qualitatively 
different from the data.   

\begin{deluxetable}{llllll}
\tablewidth{0pt}
\tabletypesize{\scriptsize}
\centering
\tablecaption{Basic instrumental parameters for the \textsc{UFig} image simulations. 
\label{tab:ufig_params}}

\tablehead{\colhead{Filter} & \colhead{$g$} & \colhead{$r$} & \colhead{$i$} & \colhead{$z$} & \colhead{$Y$}} 

\startdata
Atmosphere throughput                &   0.8           & 0.9           & 0.9        & 0.9         & 0.95          \\
Telescope throughput                 & 0.43         &  0.51       &  0.56      &   0.56     & 0.19             \\
Mean wavelength (nm)               & 473         & 638      & 775     & 922     & 995         \\
Bandwidth (nm)                          & 147           & 141         & 147        & 147        & 50               \\
Quantum efficiency                    & 0.7            & 0.75         & 0.85       & 0.8         & 0.3            
\enddata
\end{deluxetable}
     
\subsubsection{\textsc{UFig} image $\rightarrow$ DESDM catalog}
\label{sec:im2cat}
In this step we run the DESDM software on the \textsc{UFig} images to produce \textsc{SExtractor} catalogs. 
First, the PSF model is estimated by \textsc{PSFEx} on each of the single-band coadd images. Then we follow the 
procedure implemented in DESDM and make a deep ``detection image'' by stacking the coadd images in 
three bands ($riz$). Objects are detected on the ``detection image'' but the properties of each object 
are measured on the single-band images using \textsc{SExtractor}. The software versions used in this 
work are: \textsc{SExtractor} v2.18.10, \textsc{PSFEx} v3.17.0 and \textsc{SWARP} v2.36.2. The 
configuration files for \textsc{SExtractor} and \textsc{PSFEx} can be found at: 
\url{http://www.phys.ethz.ch/~ast/cosmo/bcc_ufig_public/bcc_ufig_config.tar.gz}

This is the most time-consuming step in the framework, as \textsc{SExtractor} carries out a large 
number of measurements and galaxy profile-fitting operations. However, depending on the specific science 
interest, it is possible to eliminate some of the \textsc{SExtractor} functionalities and make this step faster.
For instance, eliminating the process of fitting galaxy profiles speeds up the procedure by a factor of $\sim 100$.

\begin{figure*}
  \begin{center}
  \includegraphics[scale=0.5]{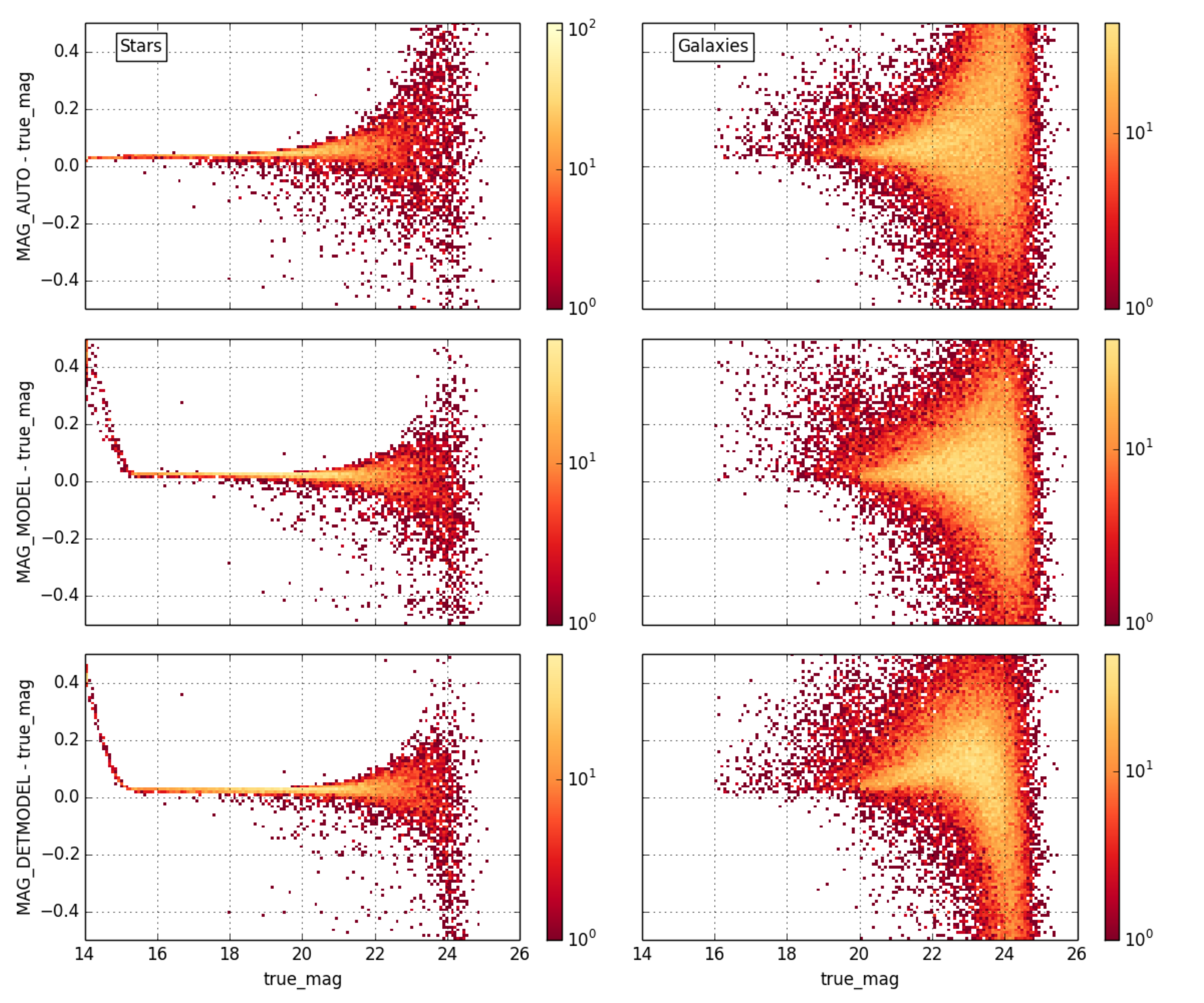}  
  \end{center}
  \caption{Distribution of the differences in three magnitude measurements and the true input magnitude 
  as a function of the input magnitude. From top to bottom are the \textsc{SExtractor} magnitudes \magauto{}, 
  \magmodel{} and \magdetmodel{}. Left and right panels are for stars and galaxies respectively. All plots are 
  generated for one arbitrary $i$-band image in our simulation. Note that the colour scales are logarithmic. }
\label{fig:mag_io}
\end{figure*}

\subsubsection{DESDM catalog $\rightarrow$ BCC catalog}
\label{sec:cat2cat}

Finally, to close the loop, the catalogs generated from \textsc{SExtractor} above are matched to the input BCC 
catalogs by the position on the sky, and a matching file containing the galaxy ID's in the input and output catalog 
is written out. \chihway{The matching process is sped up by first dividing each image into 20 smaller areas, and then 
matching within the subareas.} It is this matching that gives us a model of the transfer function for DES data. We 
now have a mapping between the input signal from the sky and the final catalogs one uses for science.  

\begin{figure*}
  \begin{center}
  \includegraphics[scale=0.5]{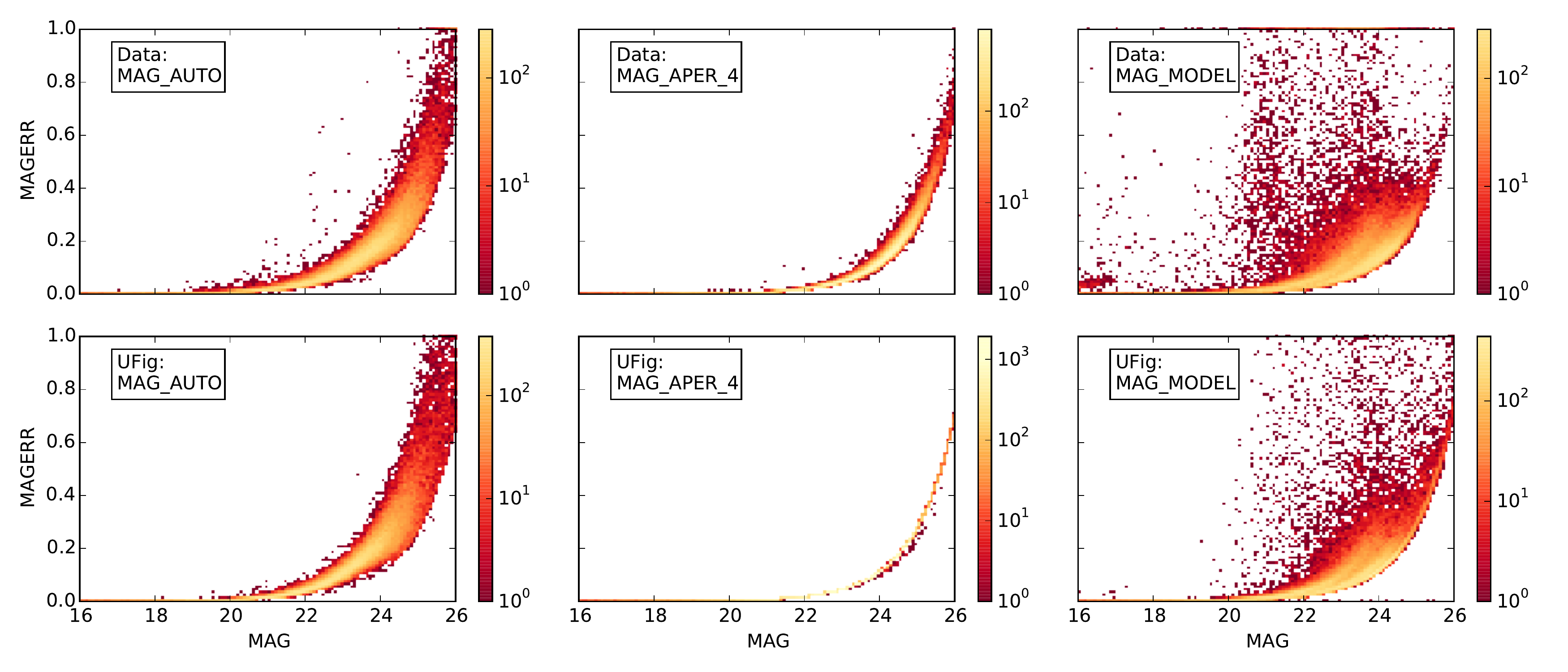}  
  \end{center}
  \caption{Distribution of of three magnitude measurements and the associated errors as quoted from the 
  \textsc{SExtractor} output. From left to right are the \textsc{SExtractor} magnitudes \magauto{}, \magaper{} 
  (2 arcsec) and \magmodel{}. The top row shows that measured from one arbitrary $i$-band SV image and the 
  bottom shows the measurement from the corresponding simulated image. The colour scales are logarithmic. 
  Note that the middle bottom panel shows that most of the data points lie on a very tight line in this parameter 
  space.}
\label{fig:mag_io2}
\end{figure*}

\subsection{Data volume and computational cost}
\label{sec:computational}

The images and catalogs in this work are generated on the Brutus cluster at 
ETH Zurich.
The typical run time to generate the FITS image and \textsc{SExtractor} catalog for a 0.75$\times$0.75 deg$^{2}$ patch 
of sky in one filter band for our SVA1 simulation set (see \Sref{sec:qa}) is summarised in \Tref{tab:computational}, together 
with the file sizes. The runtimes is calculated for running with one core on AMD Opteron 6174/8380/8384 machines.
Generally, the run time of the image generation scales with the number of photons, or exposure time and the run time 
for the analysis processes scale with the number of objects detected. The run time is dominated by the Source 
Extractor analysis process. 

Note that \Tref{tab:computational} does not include the generation of the BCC catalogs upstream to this 
work, which includes the N-body simulations and the input galaxy/star/quasar catalogs. To estimate the computational 
cost for the full end-to-end framework, one would also need to take into account these factors, which adds a total of 
$\sim 340$k CPU hours to the computational time.

\begin{deluxetable}{llll}
\tablewidth{0pt}
\tabletypesize{\scriptsize}
\centering
\tablecaption{Summary for the average runtime on one core and size of output files for the SVA1 simulations 
in this work. All numbers are quoted for one coadd image in one filter, and all data size are quoted after gzip 
compression. \label{tab:computational}}
\tablehead{\colhead{Output} & \colhead{Run time} & \colhead{Format} & \colhead{Size} } 
\startdata
Coadd image &  7.0 min & FITS & 356 M  \\
\textsc{SExtractor} catalog &  2.5 hr &FITS & 53 M  \\
Matching file & 3.8 min  &ASCII & 1.4 M
\enddata
\end{deluxetable}

\section{Quality Assurance: forward-modelling the DES SVA1 data}
\label{sec:qa}

In this section we present several basic quality assurance tests on the output catalog of the above simulation 
framework. The main goal is to show that our framework produces reliable catalogs that can be used 
for interpreting scientific data under well understood assumptions. For regimes where the simulations do not 
properly model the data, we identify areas for improvement in our model. 

We set our target to model the DES SVA1 dataset described in \Sref{sec:data}. We generate coadd images 
and catalogs covering the SVA1 footprint (\Fref{fig:footprint}) in all 5 filter bands. In addition to the basic 
parameters listed in \Tref{tab:ufig_params}, we also use compiled maps for mean observational parameters 
from the data themselves (seeing, limiting magnitude, magnitude zeropoint). These maps are generated 
similar to the systematics maps described in \citet{2013MNRAS.435.1857L}. For each of our images in each 
filter band, we find the corresponding region of sky in the maps. Then, we take the median value of the maps 
to be the observational parameters for this image. 
Note that for modelling another dataset, 
even with the same instrument, the results could differ significantly. A portion of the SVA1 simulation output and 
supporting documentation can be found at \url{http://www.phys.ethz.ch/~ast/cosmo/bcc_ufig_public/}. The total 
number of coadd images is 480 in $griz$-bands and 432 in $Y$-band.

Below we focus on examining three basic measurements of the detected objects in the images -- magnitude, 
size and object number counts.

\subsection{Magnitude}

Photometry lies at the centre of many science analyses. Yet, in typical astronomical data, 
magnitude measurements and the corresponding errors are often hard to predict from first principles due 
to the noisiness of the data, the non-linear nature of the measurement procedure, and the coupling to the 
objects' size and profile. We examine here the relation between the input and different measured magnitudes. 
Then we compare the general behaviour of the different magnitude measurements in the SVA1 data 
compared with that in our simulations. Similar analyses have been done in 
\citet{2011arXiv1109.6741S, 2011AJ....141..185R} for early DES simulations. 

In \Fref{fig:mag_io} we show the distribution of the difference between measured and input magnitude 
as a function of input magnitude for three different magnitude estimates from \textsc{SExtractor} 
(\magauto{}, \magmodel{} and \magdetmodel{}) on one arbitrarily selected $i$-band image. 
\magauto{} is measured by summing the flux in an ellipse scaled to the Kron radius 
\citep{1980ApJS...43..305K}; \magmodel{} is measured by fitting the object with a given model and 
estimating the flux for this model; \magdetmodel{} is similar to \magmodel{} but first carries out the 
model fitting on the detection image, and then fits the overall normalisation of this model to each 
single-band image separately. \magdetmodel{} thus has a consistent galaxy model for the same 
galaxy across all filters, which is primarily useful for colour measurements. For SVA1, \magmodel{} 
and \magdetmodel{} use a single exponential profile for the galaxy model.   

The general trend between all three estimates is that the measured magnitudes tend to be biased high 
and that faint objects have larger photometric errors than bright objects. The bias is due to the fact that 
the magnitudes are all calculated within some finite pixels defined by the signal-to-noise of each pixel, 
whereas in reality, light can fall much further out. For the stars, the bias is at the 0.01--0.02 level at the 
bright end, with \magauto{} slightly higher than the other two. This is sensible as the fitting methods 
(\magmodel{} and \magdetmodel{}) does account for some of the low-level wings. Model fitting 
also results in smaller scatter at the faint end and the sharp turnoff at the very bright end, where the 
model fails to fit bright star profiles. For galaxies, there is a small ``bump'' feature at magnitude $\sim20$. 
The feature is a result of the input galaxy model, where galaxies have different distribution of profiles 
above and below $i=20$ (\Eref{eq:sersic}). The galaxy \magauto{} measurements behave 
similar to that for the stars with slightly more scatter. \magmodel{} and \magdetmodel{}, however, 
does not improve significantly the magnitude measurements compared to \magauto{}. This could 
indicate that the model for the galaxy profiles used by \magmodel{} and \magdetmodel{} is insufficient 
for the wide range of galaxy profiles in the simulations (and in data). We also see 
that \magmodel{} is less biased compared to \magdetmodel{}. This is because \magdetmodel{} 
derives the galaxy model from the detection image ($riz$-coadd) instead of the image where the 
magnitude is measured. Note that, the difference would be larger in real data, where unlike in our 
simulations, the galaxy and the PSF profiles change in different filter bands.
  
\chihway{In \Fref{fig:mag_io2} we show the magnitude error against magnitude for one arbitrary $i$-band DES image 
and the corresponding \textsc{UFig} simulation. We examine the behaviour of three different magnitude estimates in 
the \textsc{SExtractor} catalog. All objects in both catalogs are plotted. The broad features in the different panels 
agree between the simulation and the data with some discrepancies that are expected from the simplifications 
and assumptions described in \Sref{sec:simulations}. First, in the \magerrauto{} - \magauto{} panel agree down to 
$i \sim 24.5$, but there are more objects in the simulations compared to the data at $i>24.5$. This shows that the 
simulation is able to reproduce the behaviour of the magnitude error at $i<24.5$, which is sufficiently deep for DES. 
For the fainter objects, one should take caution when interpreting results from the simulations in this regime. Second, 
the \magerraper{} - \magaper{} relation in the simulation lies on top of that from data. This confirms that our noise 
model behaves as expected (see \Sref{sec:noise}). The data contains more scatter compared to the simulations.   
This is expected as the limiting magnitude varies within an image in data, while we have assumed it to be 
constant in our simulations. Finally, for the \magmodel{} - \magerrmodel{} panel, both data and simulation show 
an overall more complicated shape of the distribution. The same qualitative feature can be seen in both plots, 
such as the sharp drop of numbers at \magerrmodel{}$\sim0.2$, the faint could of objects with large \magerrmodel{} 
at \magmodel{}$\sim 24$. These indicate that our model of the intrinsic galaxy morphology (size and S\'{e}rsic index) is 
reasonable. The details in the two distributions are however difference. This is an indication that improvements are 
needed in the future in this area, and one should take caution when using \magmodel{} in our simulations.}

\begin{figure*}
  \begin{center}
  \includegraphics[scale=0.5]{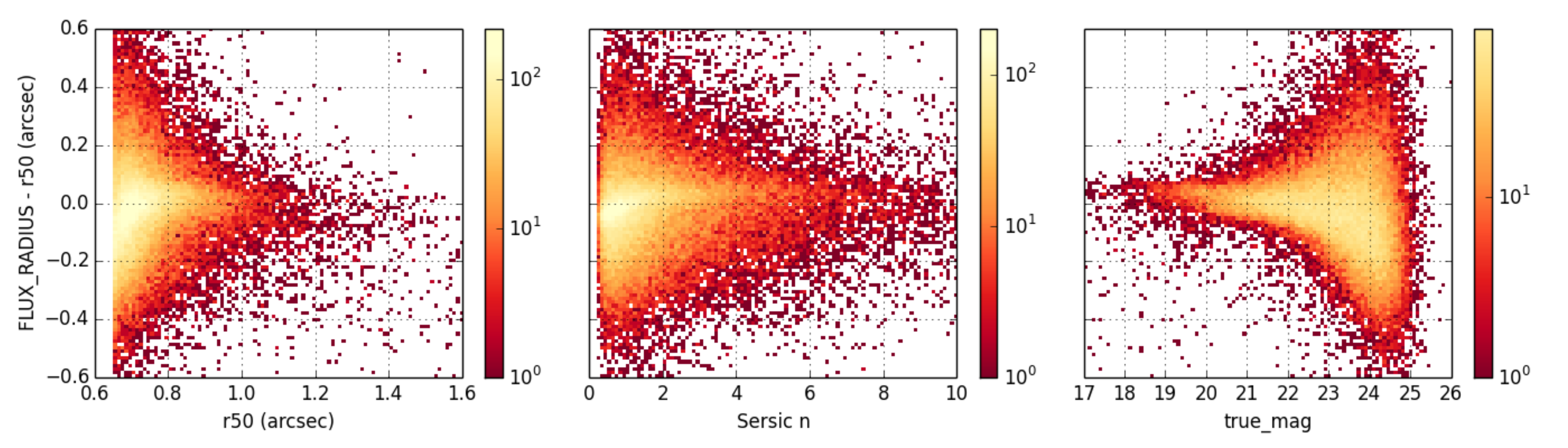}  
  \end{center}
  \caption{Distribution of the difference in measured size and input size r50 as a function of r50 (left), 
  S\'{e}rsic index (middle) and magnitude (right). r50 is defined in \Eref{eq:r50}. All plots are generated 
  for one arbitrary $i$-band image in our simulation. And more that the colour scales are logarithmic.
   }
\label{fig:size_io}
\end{figure*}

\begin{figure}
  \begin{center}
  \includegraphics[scale=0.5]{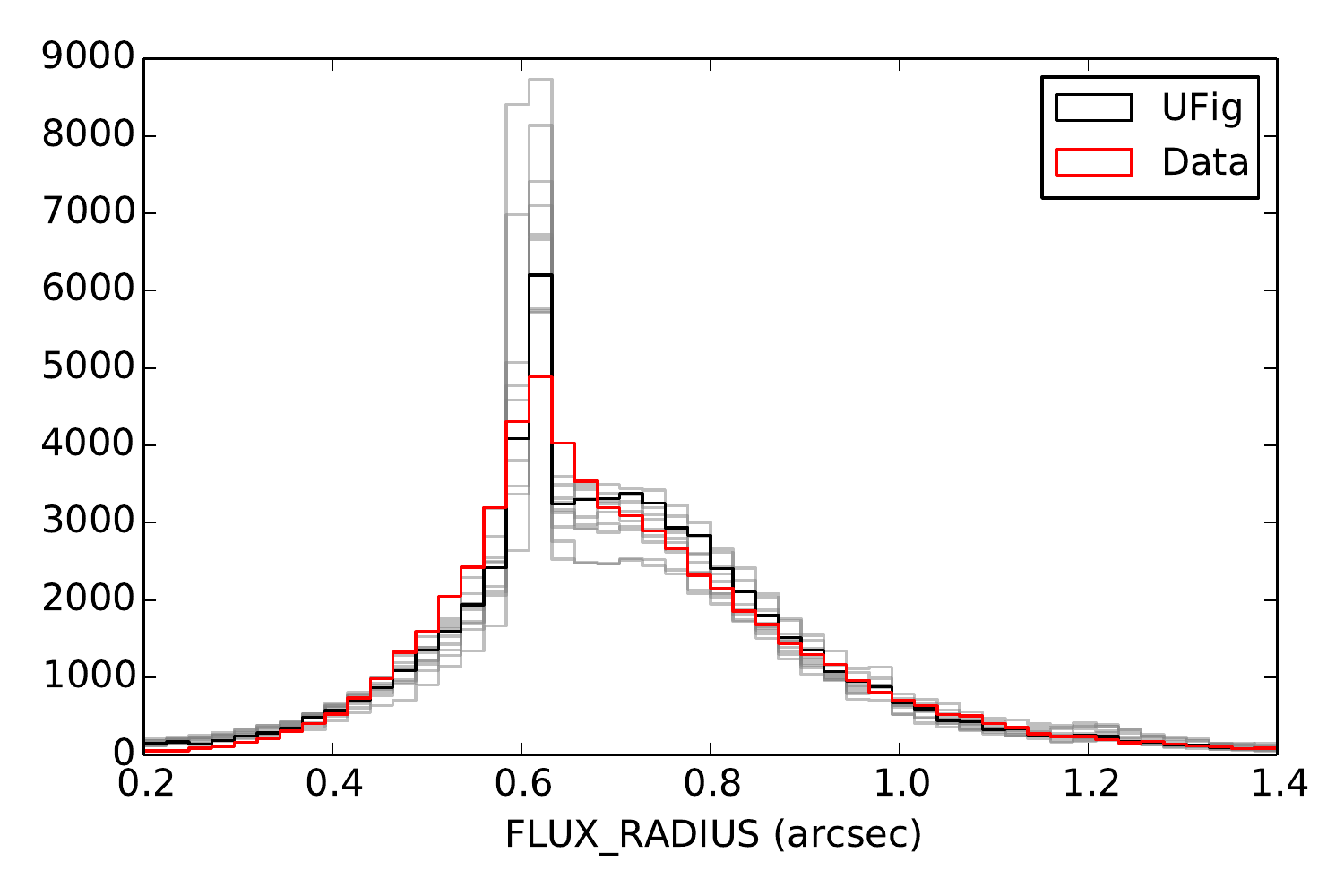}  
  \end{center}
  \caption{Measured size distribution for all objects from the \textsc{UFig} simulations (black) compared to the 
  SVA1 data (red) in the same area. 
    The grey lines show the same distribution as the black line, but for other tiles in our SVA1 simulation 
  that have limiting magnitudes and seeing conditions within 1\% of the region of interest. The disagreement 
  in the distributions is consistent with the variation from cosmic variance. }
\label{fig:size_data}
\end{figure}

\subsection{Size}

The first-order morphological information we can measure from an object's image is its observed size. The 
measured \textit{size} of an object in a noisy image is usually defined in terms of the flux in a set of pixels that are 
assigned to this object -- for example, the parameter \fluxradius \hspace{0.01in} in \textsc{SExtractor} refers to the radius within 
which 50\% of the total flux is enclosed. The measured size is thus coupled with magnitude 
measurements and is sensitive to the noise in the image, the PSF and the intrinsic object profile. 

In \Fref{fig:size_io}, we show the distribution of the difference between measured object size and input 
size (r50) as a function of input size, S\'{e}rsic index and true magnitude for all detected objects in 
one arbitrary $i$-band image. 
The ``input size'' r50 here refers to the expected half-light 
radius of the object after convolving with the PSF. We calculate it via the following empirical relation:
\begin{equation}
\rm{r50} = \sqrt{\rm{r50_{in}}^{2} + \rm{r_{PSF}}^{2}/2.355},
\label{eq:r50}
\end{equation}
where $\rm{r50_{in}}$ is the intrinsic half-light radius given by the BCC catalog and $\rm{r_{PSF}}$ is the 
seeing for that image. The numerical factor 2.355 is derived empirically to account for the change of the 
apparent galaxy size when convolved with the PSF. Note that \Eref{eq:r50} is only an approximate relation 
between $\rm{r50_{in}}$ and $\rm{r50}$. Nevertheless, we use it here to illustrate the qualitative behaviour 
of the size measurements in our catalogs.
 
\Fref{fig:size_io} shows that small, faint, disk-like galaxies have larger errors on the size measurement.
The distribution of the errors are asymmetric with more objects biased small. The origin of the asymmetry 
comes from the fact that \textsc{SExtractor} measures the sizes with a finite set of pixels while the galaxy 
profile generally extends beyond that.  

In \Fref{fig:size_data} we compare the measured size distribution of all the detected objects in one arbitrary 
$i$-band image in the SVA1 data and the corresponding simulation. Also overlaid in grey are 10 other size 
distributions from the simulations that have limiting magnitude and seeing values within 1\% of this image, 
these curves give an estimate of the variation in the size distribution due to cosmic variance. 
\chihway{We find that 
the measured size distribution in our simulations are consistent with that measured in data within cosmic 
variance. The narrow peak at \fluxradius$\sim0.6$ arcsec corresponds to the seeing value for this image. 
The peak is broadened in the data since unlike in the simulations, there exists 
seeing variation within each image. The size distribution of the remaining objects (mostly galaxies) match 
very well between the data and simulations, especially on the high and low end where it is less sensitive to 
our assumption of constant seeing. Seeing variation is thus one important factor to improve in future 
developments. }  

\subsection{Number density}

 \begin{figure*}
  \begin{center}
  \includegraphics[scale=0.5]{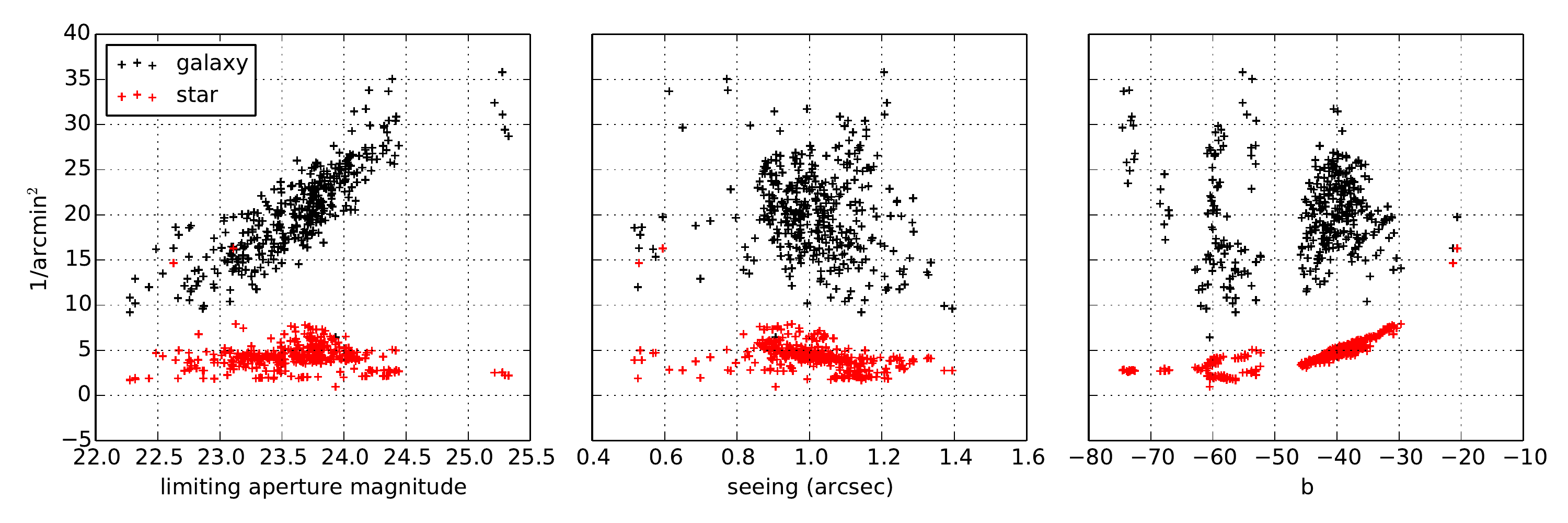}  
  \end{center}
  \caption{Galaxy (black) and star (red) number densities as a function of the limiting (2 arcsec) aperture magnitude 
  (left), seeing (middle) and galactic latitude (right). These numbers are calculated from all objects detected in all the 
  $i$-band images in the SVA1 simulations in this work. Each data point represents one image in our simulation. The 
  discontinuous distribution of data points in the x-axis of the right panel reflects the SVA1 footprint.}
\label{fig:number_count}
\end{figure*}

\begin{deluxetable}{lll}
\tablewidth{180pt}
\tabletypesize{\scriptsize}
\centering
\tablecaption{Object number density (per sq. arcmin) from data and our simulations under different magnitude 
(\magauto{}) cuts. \label{tab:number_count}}
\tablehead{\colhead{} & \colhead{Data} & \colhead{Simulation} } 
\startdata
All objects     & 27.79  & 31.05  \\
$15<i<19$ & 1.06 & 1.01  \\
$15<i<21$ & 3.43  & 3.85 \\
$15<i<23$ & 11.95  & 12.82
\enddata
\end{deluxetable}

Finally, we examine the detected star and galaxy number densities. This is important because it 
simultaneously checks the input source distribution, the image simulation and the analysis software.

In \Fref{fig:number_count} we show the star and galaxy number density in all the $i$-band simulated SVA1 
images as a function of limiting magnitude, seeing and galactic latitude. We observe that the general 
behaviour of the number counts follows expectation. In deeper fields the number density of stars and 
galaxies both increase. The group of data points on the far right are the Supernova fields (see 
\Fref{fig:footprint}) where the total exposure time is significantly longer than in the rest of the fields. Note 
however, the input BCC catalogs are not necessarily complete at those magnitudes, thus one should be 
careful in interpreting the results there and only treat those data points as lower bounds. The dependence 
on seeing is also expected (keeping in mind that seeing and limiting magnitude are not independent) -- higher 
seeing gives slightly lower number density since the signal-to-noise of the objects decreases going to higher 
seeing. Finally, we look at the correlation between number density and the galactic latitude as a check for the 
input source catalog. We find that the stellar density, as expected, increases towards the galactic plane, 
whereas the galaxies do not. The discontinuous distribution of data points in the x-axis reflects the SVA1 
footprint. 

To compare the number counts derived from simulations and data, we calculate the mean source density 
as a function of magnitude cuts for both the SVA1 catalog and our simulations. \chihway{We use all objects 
in the catalogs and do not make distinction between stars and galaxies. We choose to do so to avoid making 
choices in the object selection. This also means that we are accounting for spurious detections from noise, 
blended objects and artefacts. \Tref{tab:number_count} summarises our results. We find that the data and 
the simulations agree at the $\sim$10\% level. The agreement is best at the bright end, where the errors in 
the object property as well as the noise is more accurate. The agreement is not perfect, but rather 
encouraging, given the current uncertainty in the source catalog, the galaxy profile model and the noise 
model.} 

\section{Applications}
\label{sec:applications}

In this section we describe two example cases where we use the simulation products described in \Sref{sec:qa} 
to help answer questions in the data analysis process. The advantage of using this framework is 
that the simulations are sufficiently realistic, yet, we have full control over every stage of the simulation 
and data processing pipeline. For use of our simulations in scientific analyses on the DES SV 
data, see Rykoff et al. (in prep.). 

\begin{figure*}
  \begin{center}
  \includegraphics[scale=0.5]{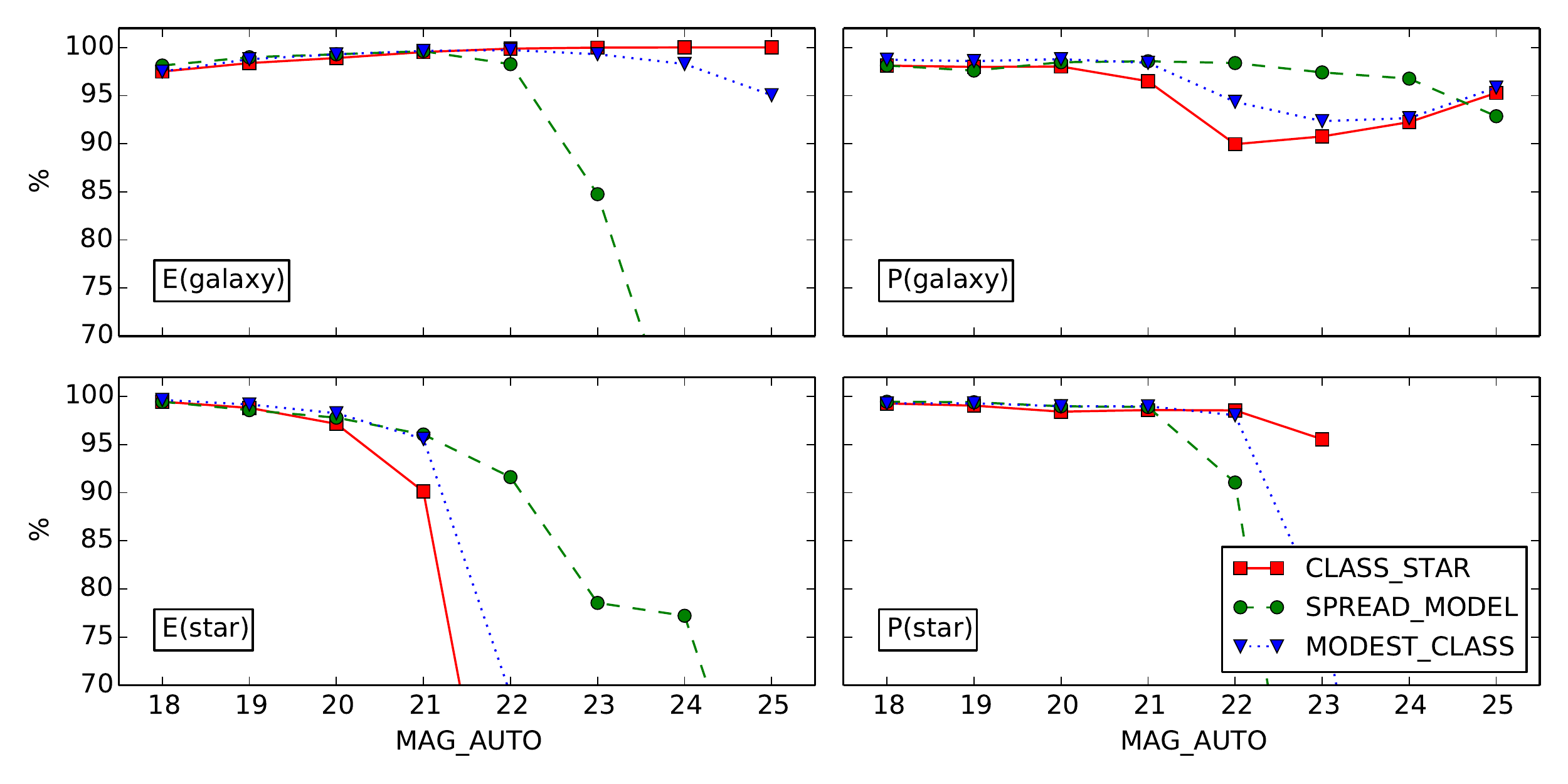}  
  \end{center}
  \caption{Efficiency (E) and purity (P) for the star and galaxy sample of three star-galaxy classifiers for 
  one arbitrary $i$-band images in our SVA1 simulations. The three classifiers are described in \Tref{tab:sg_cuts}.}
\label{fig:qa_sg_i}
\end{figure*}

\begin{deluxetable}{ll}
\tablewidth{200pt}
\tabletypesize{\scriptsize}
\centering
\tablecaption{Cuts used in the three classifiers: \classstar, \spreadmodel{} and \modestclass. 
All of these cuts have an additional cut on \flag{}$<=$3 and 5 sigma detection. For full description of 
\modestclass{} see footnote below. 
\label{tab:sg_cuts}}
\tablehead{ \colhead{Galaxies} & \colhead{Stars} } 
\startdata
\classstar$<$0.95 & \classstar$>$0.95   \\
\spreadmodel$>0.002$ & \spreadmodel$<0.002$  \\
\modestclass{} =1$^{a}$  & \modestclass{}=2$^{b}$
\enddata
\tablenotetext{a}{\modestclass=1: ({\tt FLAGS} $<=$3) AND ( NOT (\classstar{} $>$ 0.3) AND 
(\magauto{} $<$ 18.0) OR ((\spreadmodel + 3*{\tt SPREADERR\_MODEL}) $<$ 0.003) OR
(({\tt MAG\_PSF} $>$ 30.0) 
AND (\magauto{} $<$ 21.0)))}
\tablenotetext{b}{\modestclass=2: ({\tt FLAGS} $<=$3) AND ((\classstar{} $>$ 0.3) 
AND (\magauto{} $<$ 18.0) AND ({\tt MAG\_PSF} $<$ 30.0) OR (((\spreadmodel +
3*{\tt SPREADERR\_MODEL}) 
$<$ 0.003) AND ((\spreadmodel +3*{\tt SPREADERR\_MODEL}) $>$ -0.003)))}
\end{deluxetable}

\begin{figure*}
  \begin{center}
  \includegraphics[scale=0.5]{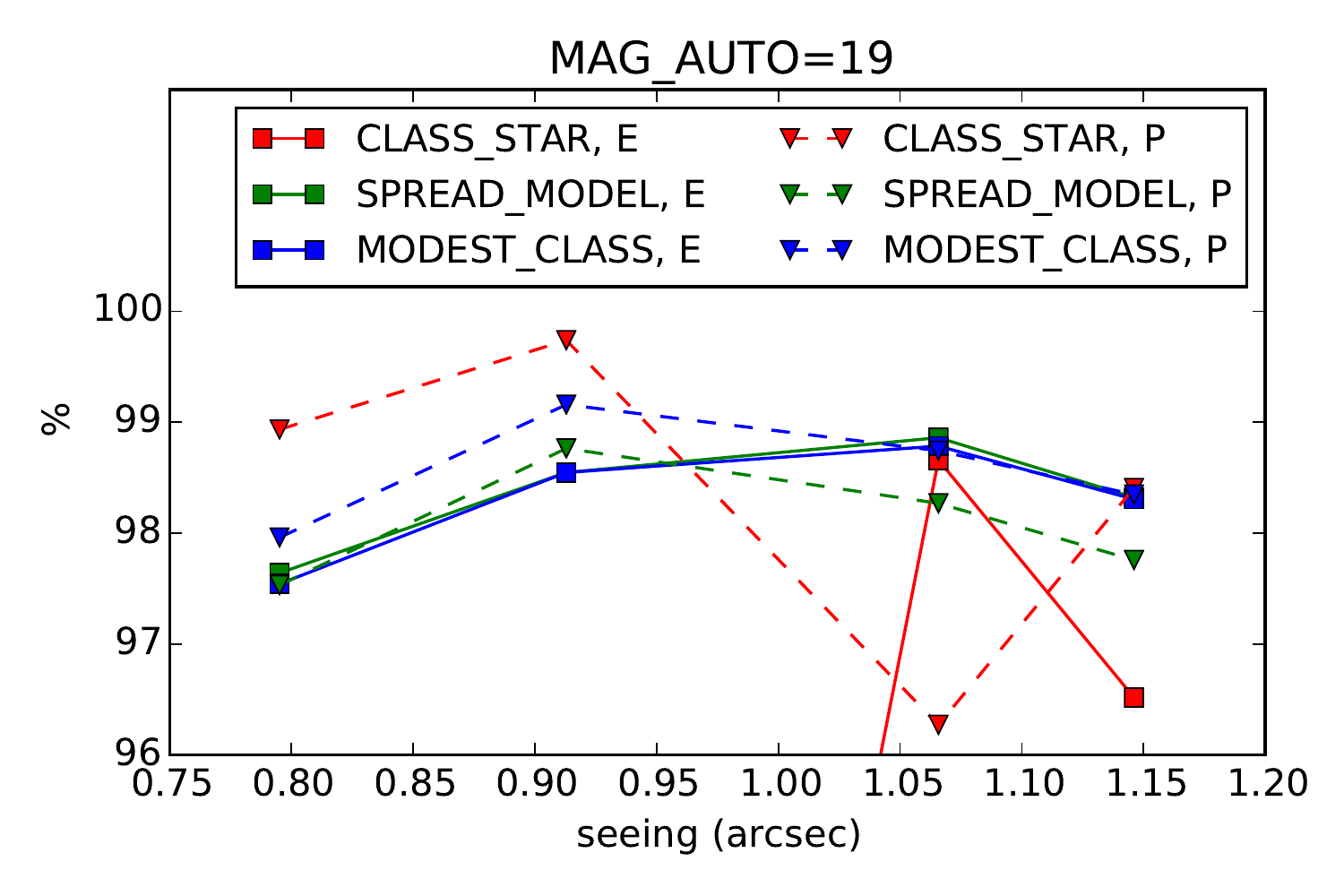}  
  \includegraphics[scale=0.5]{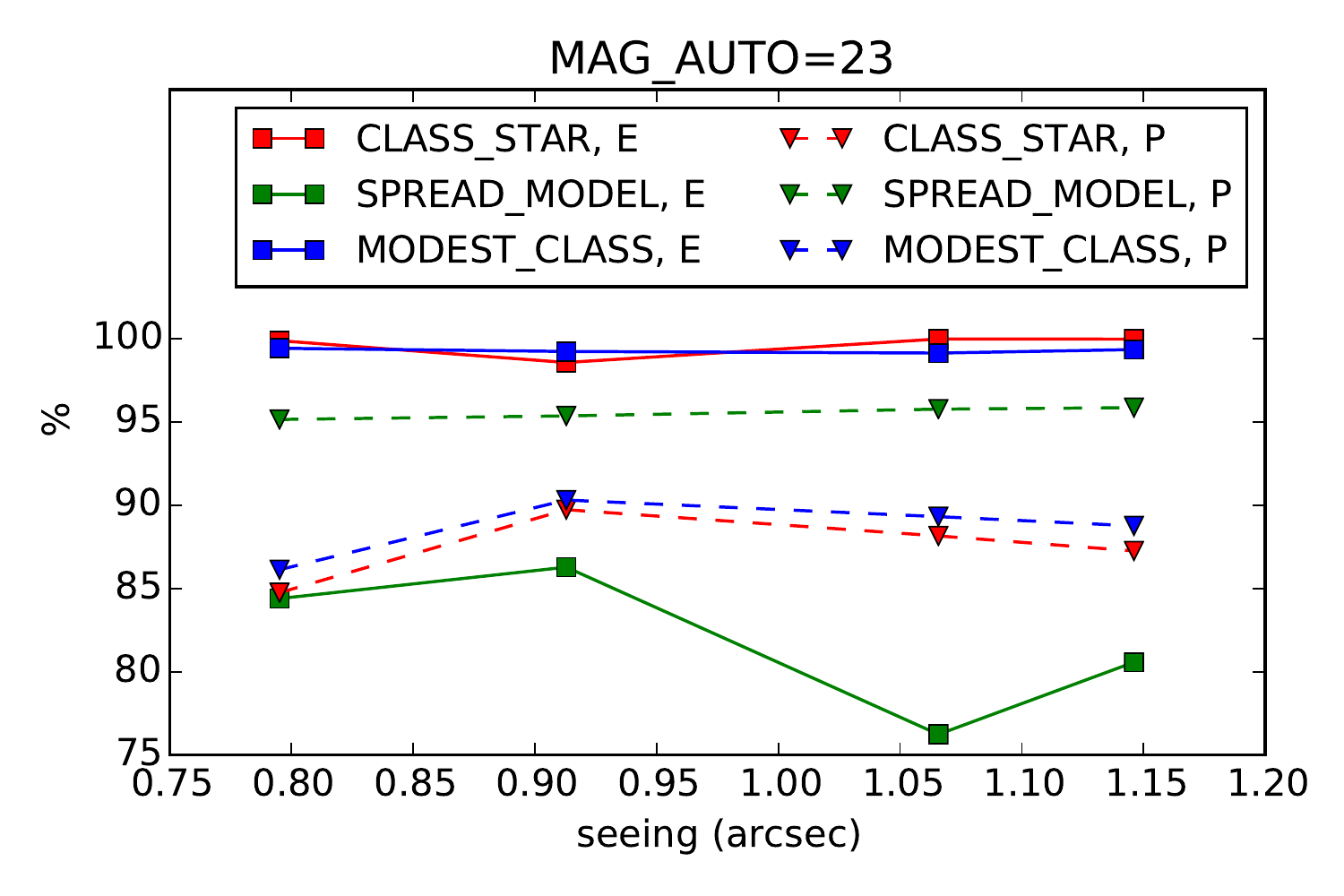}  
  \end{center}
  \caption{Median efficiency (E) and purity (P) for galaxy classification of all simulated SVA1 images at 
  $18.5<$\magauto{}$<19.5$ (left) and $22.5<$\magauto{}$<23.5$ (right), as a function of seeing of 
  that image. The three classifiers are described in \Tref{tab:sg_cuts}. Note that the y-axis has very different scales. }
\label{fig:qa_sg_i2}
\end{figure*}

\subsection{Star-galaxy classification}
\label{sec:sg}

Identifying stars and galaxies in optical images is one of the most basic operations in the data analysis pipeline. 
Depending on the science application, one would demand good efficiency and/or purity in the star sample and/or 
the galaxy sample. For example, in weak gravitational lensing, one would require a pure star sample for the PSF 
estimation, and a pure galaxy sample for un-contaminated lensing signal. On the other hand, for study of galaxy 
evolution, the completeness of the galaxy sample is also important in order for one to extract global behaviours 
of the galaxy population. We define the star/galaxy classification efficiency ($E$) and purity ($P$) below:
\begin{equation}
E ({\rm X})= \frac{{\rm \# \;of \;objects \;correctly \;identified \;as \;X}}{{\rm \# \;of \;all \;X}}
\end{equation}
\begin{equation}
P ({\rm X})= \frac{{\rm \# \;of \;objects \;correctly \;identified \;as \;X}}{{\rm \# \;of \;objects \;identified \;as \;X}}
\end{equation}
where X is either stars or galaxies. 

The problem is challenging, however, in typical ground-based imaging data. With typical seeing and noise 
conditions in these images, small, faint galaxies become indistinguishable from stars. A wide range of techniques 
have been developed to resolve this problem \citep{2011MNRAS.412.2286H, 2012ApJ...760...15F, 
2013arXiv1306.5236S}. Standard star-galaxy classifiers use morphological information of the stars, more advanced 
ones incorporate also the colour information \citep{2010AA...514A...3P}. The simulations from this work, with both 
realistic image characteristics and colour information, offer a generic tool for different methods to be tested on 
before applying to data. Moreover, since the simulations are tailored for a specific set of data, one can 
consistently evaluate the effect of star-galaxy separation on specific science measurements performed on 
the same dataset. 

\begin{figure*}
  \begin{center}
  \includegraphics[scale=0.5]{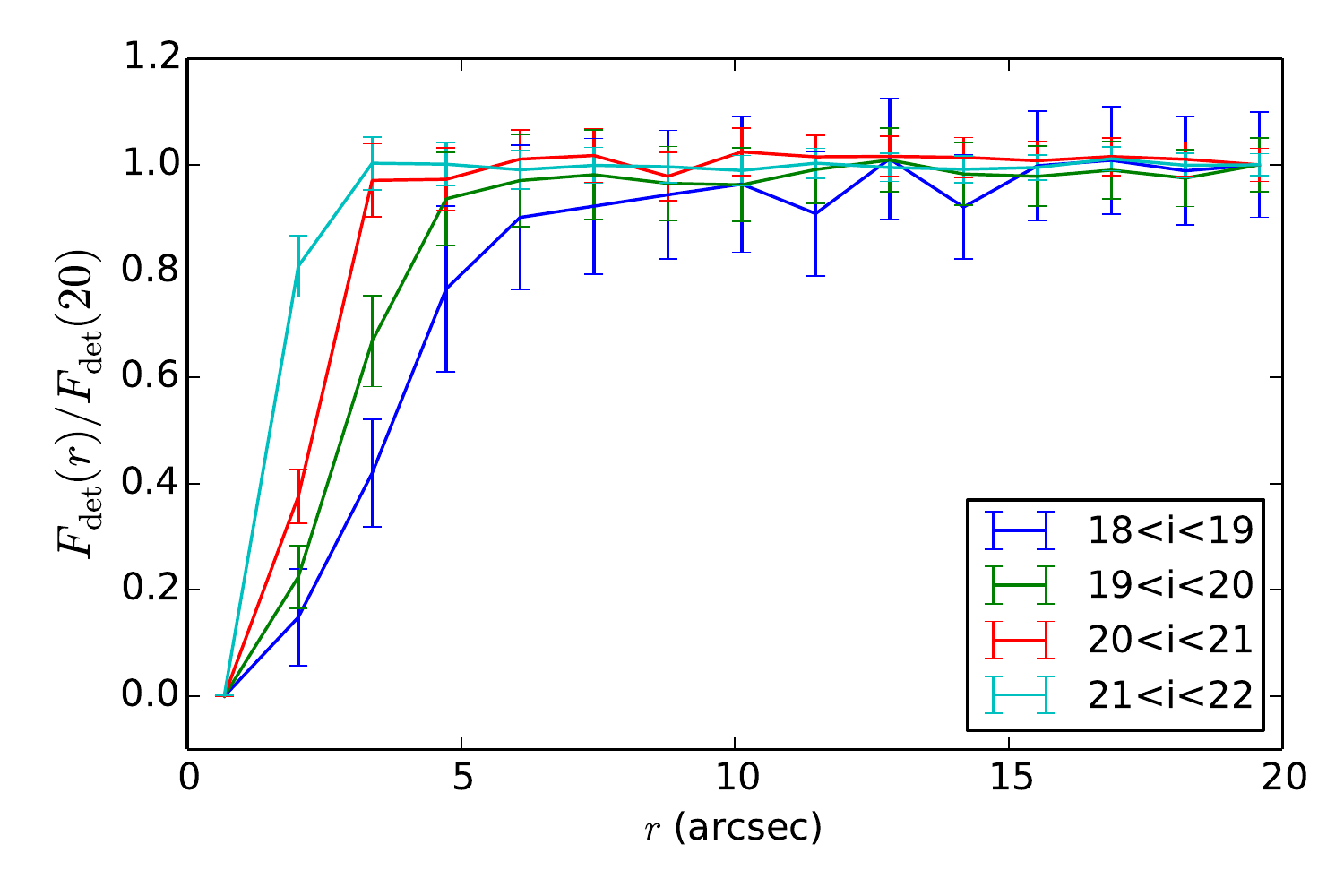}  
  \includegraphics[scale=0.5]{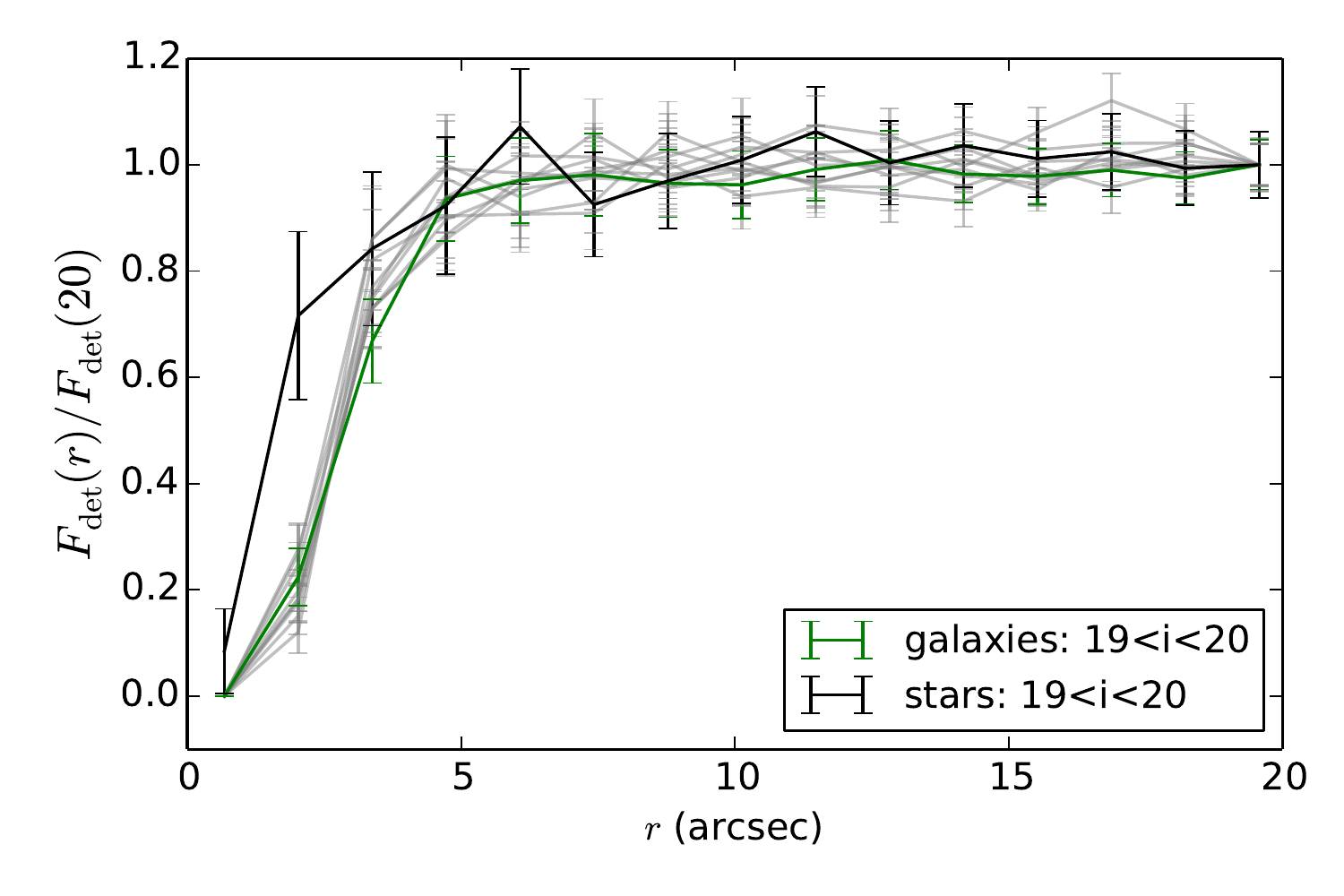}  \\
  \end{center}
  \caption{Degradation of detection efficiency due to proximity effects around bright objects evaluated for 
  one arbitrary $i$-band image in our SVA1 simulation. In the left panel, the four curves indicate the detection 
  efficiency for source galaxies in the magnitude range $18<i<24$ around center galaxies in different magnitude 
  bins. The x-axis shows the distance from the center galaxy. The y-axis shows the fraction of source galaxies 
  detected. All curves are normalized so that the level measured at 20 arcsec is 1. This removes the detection 
  efficiency from the finite depth. The right panel shows only the magnitude bin $19<i<20$ in the left panel, but 
  overlays in grey results for 10 other random images. The grey lines agree with the green line within error bars, 
  despite the different observational conditions. Also plotted in black is the result when we replace the centre 
  galaxies with stars, which results in a qualitatively different shape of curve.}
\label{fig:proximity}
\end{figure*}

Here, we show an example of quantifying the performance of three single-band cut-based star-galaxy 
classifiers which are based solely on the SExtractor catalogs. The three classifiers
which we label as \classstar, \spreadmodel, and \modestclass{} are described in
\Tref{tab:sg_cuts}.  \classstar{} is a pre-trained Artificial Neural Network method that uses several of 
the photometric and shape information in the 
\textsc{SExtractor} catalogs.  It works well at the bright end but is limited by
requiring the user to know the approximate seeing of the image prior to
processing.  \spreadmodel{} \citep{2012SPIE.8451E..0DM, 2013AA...554A.101B} uses pixel-level 
morphological information and compares
the profile of each object with the local PSF.  For faint objects, where the
classification is most challenging, \classstar{} with the current settings tends to classify all objects as
galaxies at the faint end while a naive \spreadmodel{} classifier with constant threshold tends to classify 
all objects as stars. \modestclass{} is a new classifier used for SVA1 Gold that
has been developed empirically and tested on DES imaging of COSMOS fields with
Hubble Space Telescope ACS imaging.  It is primarily based on \spreadmodel, and
attempts to fix the faint galaxy classification by including the error on
\spreadmodel.  

\chihway{We evaluate the $\rm E$ and $\rm P$ statistics for stars and galaxies on one arbitrary $i$-band image in our SVA1 
simulations as a function of the measured \magauto{}. The results are shown in \Fref{fig:qa_sg_i}. In this particular 
image, the simulations confirm nicely what we expect from the construction of the three classifiers (see above). For example, 
for galaxies, \spreadmodel{} gives high $\rm P$ and low $\rm E$ at the faint end, \classstar{} behaves in the opposite 
direction, and \modestclass{} sits between the two. We also see that all classifiers perform well at the bright 
end while degrading at the faint end. }

In \Fref{fig:qa_sg_i2}, we plot the median of the $\rm E$ and $\rm P$ statistics for galaxies and for all the SVA1 simulations 
as a function of seeing. The statistics is evaluated at $18.5<$\magauto{}$<19.5$ and $22.5<$\magauto{}$<23.5$ 
to illustrate the global performance of the different classifiers at bright and faint magnitudes. We find that 
\classstar{} is unstable at the bright end $i\sim19$, while the other two perform well. At the faint end, \modestclass{} 
improves from \spreadmodel{} in $\rm E({\rm galaxy})$, consistent with \Fref{fig:qa_sg_i}. There are mild dependence on 
seeing for \spreadmodel{} and \modestclass{} at the bright end and all classifiers at the faint end. Interestingly, the galaxy 
classification purity rises going towards larger seeing and drops after $\sim1.05$ arcsec. 

As there are simplifications in both our galaxy and PSF, we do not expect these results should reproduce quantitatively 
exactly the same in data. However, the simulations allow us to study the response of different star-galaxy classifiers 
to observational parameters and object properties. Understanding the physical interpretation for their behaviours in the 
simulations then helps us quantify the contamination in our star/galaxy sample in data. 

\subsection{Proximity effects on object detection}

Object detection software for imaging data, such as \textsc{SExtractor}, relies on identifying a group of pixels 
that have values above the local background level at some predefined signal-to-noise threshold. As a result, 
the probability of detecting an object depends on the object brightness and the local pixel values around that 
object -- these pixels contain not only the sky background but also photons from other objects nearby. The 
proximity effect on object detection refers to the fact that for the same object and sky background, we are less 
likely to detect it when there exist nearby bright objects. This effect is especially pronounced in crowded 
environments such as galaxy clusters or dense stellar fields \citep{2014arXiv1405.4285M, 2014arXiv1409.2885Z}, 
but can also affect more generally the clustering statistics for large-scale structure \citep{2012MNRAS.424..564R, 
2014ApJ...780L..16H}. 

Calibrating the effect from data itself is possible, but can be coupled with other factors such as photometric 
errors and star-galaxy classification. On the other hand, simple catalog-level simulations are inefficient for 
this specific problem, as the object detection algorithm is a highly non-linear operation and needs to be 
performed on images. Image-level simulations, such as that developed in this work are ideal for this test, as 
it contains the following key features that are required to perform this analysis: (1) realistic spatial distribution 
(clustering) of galaxies and stars, (2) realistic observed magnitude distribution of stars/galaxies and 
morphology distribution for galaxies, and (3) image-level simulations that are processed through the same 
object detection software as the data. In this section, we demonstrate an example where we quantify via 
simulations the degradation in detection efficiency due to the proximity effect. The approach of using 
simulations to correct for these effects has been used in recent literature. For example, 
\citet{2014arXiv1405.4285M} used simulations from the Balrog\footnote{\url{https://github.com/emhuff/Balrog}} 
code to asses how the crowded cluster environment reduces the probability of performing weak lensing 
measurements near the centre of galaxy clusters.  

We calculate the detection efficiency $F_{\rm det}(r)$ at a distance $r$ around a particular sample of objects 
(e.g., bright galaxies). $F_{\rm det}(r)$ is defined as
\begin{equation}
F_{\rm det}(r) = \frac{\Sigma_{i}^{n} N_{i, {\rm det}}(r)}{\Sigma_{i}^{n} N_{i, {\rm true}}(r)}
\end{equation}
where $i$ is summed over the $n$ objects in this sample of interest, $N_{i, {\rm det}}(r)$ is the number of 
objects detected at a distance $r$ and $N_{i, {\rm true}}(r)$ is the true number of objects at this distance. 
Without the proximity effect, we expect the $F_{\rm det}(r)$ curve to be flat.  

In \Fref{fig:proximity} we show the $F_{\rm det}(r)$ for an arbitrary $i$-band image in our SVA1 simulations. 
Here we set up the calculation to estimate the detection efficiency of galaxies at $18<i<24$ in the surrounding 
of other galaxies in different (true) magnitude bins. For clarity, we will refer to the objects responsible for the drop 
in detection efficiency the ``center'' objects and the objects being detected the ``source'' objects. We would like to 
know how many source galaxies are missing in the magnitude range of $18<i<24$ because there is a center 
galaxy nearby. We find that the proximity effect is most severe in the surrounding of bright center galaxies, and 
the effect is seen up to several arc seconds away from the centre galaxy. In the most severe case in this test 
($18<i<19$ center galaxies), the detection of the source galaxies is 50\% less efficient at $\sim4$ arcsec. For 
comparison, the average measured galaxy size (\fluxradius) in this image is $\sim0.96$ arcsec.     

On the right panel of \Fref{fig:proximity} we only show the detection efficiency for the magnitude bin $19<i<20$, 
and overlay grey curves calculated from 10 random fields that have a range of limiting magnitude and seeing 
conditions. The grey curves agree well with the blue within error bars. This shows that neither cosmic variance 
nor seeing and limiting magnitude play a significant role in this calculation, i.e., the proximity effect is 
roughly at the same level for all galaxies in this magnitude bin across the sky under any observational conditions. 
However, if we calculate the same effect around stars in the same magnitude bin, as shown by the black curve, 
the shape of the curve changes and the detection efficiency increases at small separations. This is as expected 
since the stars have less extended profiles and are less likely to affect measurements in its surrounding pixels.    

One can imagine many more similar tests using these simulations to quantify the proximity effects as a 
function of crowding, galaxy size and profiles etc., which would be required depending on the science analysis 
of interest. We will not carry out the analyses here, but only point out via the example above that by properly 
using simulations, one can correct for the proximity effects in the data that are otherwise difficult to estimate.

\section{Conclusions}
\label{sec:summary}

Precision cosmology in ongoing and future optical surveys critically depend on the control of systematic effects. 
In this generation, end-to-end simulations will play an important role in understanding these systematic effects. 
In this paper we describe a framework for forward-modelling the transfer function for the Dark Energy Survey 
(DES) that takes the astronomical sources to realistic pixel-level data products such as images and catalogs. 
The same framework can be adjusted for other surveys and datasets.

We use the Blind Cosmology Challenge (BCC) catalogs as the source of astronomical objects, and simulate 
realistic images using the Ultra Fast Image Generator (\textsc{UFig}). We then perform image analysis to output 
catalog-level products. \chihway{We demonstrate the usage of this framework by forward modelling the early 
Science Verification (SV) data products from DES. We design the simulations and the analysis procedure to mimic 
closely that of the SV data, and show that our simulations reproduce many major characteristics of the data. There 
are small differences between the data and the simulations in certain areas of parameter spaces (e.g. small faint 
objects), but they can be explained by our simplified models and do not affect significantly the usage 
of the simulation as long as one is aware of the simplifications. By connecting the output measurement back to 
the input object-by-object, we have a powerful tool to investigate data-related systematic issues. We present two 
examples of such usage looking at star-galaxy classification and proximity effects.}

This is the first implementation of such end-to-end simulation efforts for ongoing large optical surveys. In the 
process we have made simplifications that we understand and will improve on continuing into future work. These 
include (1) more sophisticated models for the source morphological distribution (2) more realistic and spatially 
varying models for the PSF and the background and (3) extending the current framework to also model the 
single-exposure images and the coadd procedure. This constantly developing simulation framework that forward 
models the data side-by-side as DES continues to release data, provides a powerful tool to understand and 
interpret data in a clean and controlled fashion. The concept can also be extended to future surveys, where the 
need to understand details in the data products is even more demanding. \\

\acknowledgments{\textit{Acknowledgement}} 
We are grateful for the extraordinary contributions of our CTIO colleagues 
and the DES Camera, Commissioning and Science Verification teams in 
achieving the excellent instrument and telescope conditions that have made 
this work possible. The success of this project also relies critically on the 
expertise and dedication of the DES Data Management organization.

We thank Gary Bernstein, Eric Huff, Tesla Jeltema, Huan Lin and Felipe 
Menanteau for helpful comments and discussions on the paper. CC, AR, AA 
and CB are supported by the Swiss National 
Science Foundation grants 200021-149442 and 200021-143906. MTB, RHW, ER 
and MRB acknowledge support from the Department of Energy 
contract to SLAC National Accelerator Laboratory no. DE-AC3-76SF00515.
BL is supported by the Perren Fund and the IMPACT Fund. 
HVP is supported by STFC and the European Research Council under the 
European Community's Seventh Framework Programme (FP7/2007- 2013) / 
ERC grant agreement no 306478-CosmicDawn.  ACR is supported by the 
PROGRAMA DE APOIO AO POS-DOUTORADO NO ESTADO DO RIO DE 
JANEIRO - PAPDRJ. DG was supported by SFB-Transregio 33'The Dark 
Universe' by the Deutsche Forschungsgemeinschaft (DFG) and the DFG cluster 
of excellence 'Origin and Structure of the Universe'. 
AP is supported by DOE grant DE-AC02-98CH10886. 
JZ acknowledges support from the European Research Council in the form of a 
Starting Grant with number 240672.

Funding for the DES Projects has been provided by the U.S. Department
of Energy, the U.S. National Science Foundation, the Ministry of
Science and Education of Spain, the Science and Technology Facilities
Council of the United Kingdom, the Higher Education Funding Council
for England, the National Center for Supercomputing Applications at
the University of Illinois at Urbana-Champaign, the Kavli Institute of
Cosmological Physics at the University of Chicago, Financiadora de
Estudos e Projetos, Funda{\c c}{\~a}o Carlos Chagas Filho de Amparo
{\`a} Pesquisa do Estado do Rio de Janeiro, Conselho Nacional de
Desenvolvimento Cient{\'i}fico e Tecnol{\'o}gico and the
Minist{\'e}rio da Ci{\^e}ncia e Tecnologia, the Deutsche
Forschungsgemeinschaft and the Collaborating Institutions in the Dark
Energy Survey.

The Collaborating Institutions are Argonne National Laboratory, the
University of California at Santa Cruz, the University of Cambridge,
Centro de Investigaciones Energeticas, Medioambientales y
Tecnologicas-Madrid, the University of Chicago, University College
London, the DES-Brazil Consortium, the Eidgen{\"o}ssische Technische
Hochschule (ETH) Z{\"u}rich, Fermi National Accelerator Laboratory,
the University of Edinburgh, the University of Illinois at
Urbana-Champaign, the Institut de Ciencies de l'Espai (IEEC/CSIC), the
Institut de Fisica d'Altes Energies, Lawrence Berkeley National
Laboratory, the Ludwig-Maximilians Universit{\"a}t and the associated
Excellence Cluster Universe, the University of Michigan, the National
Optical Astronomy Observatory, the University of Nottingham, The Ohio
State University, the University of Pennsylvania, the University of
Portsmouth, SLAC National Accelerator Laboratory, Stanford University,
the University of Sussex, and Texas A\&M University.

This paper has gone through internal review by the DES collaboration.


\appendix

\section{A. Noise level in \textsc{UFig} images}
\label{sec:noise}

The noise level in images affects object detection, photometry measurements, and the completeness of 
the final catalog. As a result, we want to simulate images with noise properties as close as possible to that 
of the data. However, characterising the background level in the data is itself a challenging task, let alone 
the fact that we wish to model the effect of the background noise with just a simple constant Gaussian noise.    
In this work, we take an approximate approach using \textsc{SExtractor} quantities and empirically calibrate 
the noise level instead of deriving it from first principles. We defer a more sophisticated background model 
to future work. 

The basic idea is that the aperture magnitude error vs. aperture magnitude relation, for large enough apertures, 
is only a function of the background noise. Thus, once we know this 1-1 relation as a function of background 
noise, we could in principle apply the appropriate background noise level to the simulations. In principle, this 
relation could be derived analytically and the procedure described below is unnecessary. However, since our 
background model includes a Lanczos resampling, this changes slightly the statistical property of the noise, 
complicating the relation. In addition, we want to avoid any potential nonlinear processes in \textsc{SExtractor} 
that we could be missed in the calculation. 
 
Operationally, we calibrate the noise at the 10-$\sigma$ galaxy limiting (2 arcsec) aperture magnitude. That is, 
the 2 arsec aperture magnitude where the magnitude error is $\frac{2.5}{10 \ln(10)}\sim$0.1086. The calibration 
procedure is described below:  
\begin{itemize}
\item Generate \textsc{UFig} images with the median seeing of the data and a range of different background levels. 
\item Run \textsc{SExtractor} on the simulated images in the same way as on the SV data.
\item Make cuts {\tt FLAGS}==0 and \classstar$<$0.9 on the Source Etractor to get a clean sample of galaxies. 
\item Bin the galaxies in \magaper{} bins of 0.01 and find the the bin where \magerraper{}$\sim$0.1086, 
this \magaper{} corresponds roughly to the 10-$\sigma$ galaxy limiting magnitude.
\item For these simulations, plot the noise level vs. 2 arcsec aperture limiting magnitude and fit the relation. 
\end{itemize}

\begin{figure}
  \begin{center}
  \includegraphics[scale=0.55]{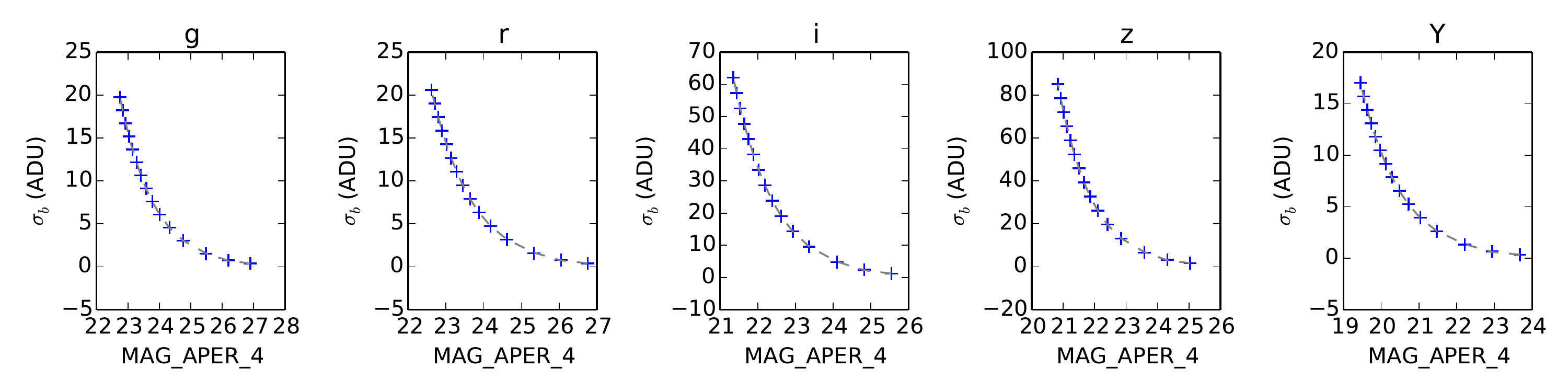}
  \end{center}
  \caption{The relation between the 2 arcsec limiting aperture magnitude and the noise level in the \textsc{UFig} images. 
  the blue points are the median of measurements in 10 random fields and the grey dashed line is the 4th-order 
  polynomial fit to these data points. \\}
\label{fig:lim_m}
\end{figure}

In \Fref{fig:lim_m}, we show the final derived calibration curve used to convert an desired aperture limiting 
magnitude to a noise level we input to \textsc{UFig}. This calibration will change slightly for images with different seeing 
and source population, but at the level of accuracy ($\sim$0.02 mag) is sufficient for our purpose here. 
 
\end{document}